\documentclass[final,3p,times,twocolumn]{elsarticle}
\pdfoutput=1 
\DeclareGraphicsExtensions{.pdf}
\usepackage{graphicx}
\usepackage{bm} 
\usepackage{atlasphysics} 
                           
\usepackage{multirow}
\usepackage{amsmath}
\usepackage{float}
\usepackage{subfigure}
\usepackage{rotating}
\usepackage{wasysym}

\newcommand{\atlasnote}[1]{\def\@atlasnote{#1}}

\newcommand{\fqqf}[1]{\ensuremath{f_{q\bar{q}}={#1}\% }}
\newcommand{\fqq}{\ensuremath{f_{q\bar{q}}}}
\newcommand{\spinzero}{\ensuremath{J^P=0^+}}
\newcommand{\spinzeromm}{\ensuremath{J^P=0^-}}
\newcommand{\spin}{\ensuremath{J^P}}
\newcommand{\spintwo}{\ensuremath{J^P=2^+}}

\newcommand{\spinalt}{\ensuremath{J^{P}_{ \mathrm {alt} }}}
\newcommand{\CLs}{\ensuremath{{\rm CL}_{\rm s}}}
\newcommand{\spinzerop}{\ensuremath{0^{+}}}
\newcommand{\spinzerom}{\ensuremath{0^{-}}}
\newcommand{\spinonep}{\ensuremath{1^{+}}}
\newcommand{\spinonem}{\ensuremath{1^{-}}}
\newcommand{\spintwopm}{\ensuremath{2^{+}_{m}}}
\newcommand{\spintwop}{\ensuremath{2^{+}}}

\newcommand{\Zjets}{$Z$+jets}
\newcommand{\mll}{\ensuremath{m_{\ell\ell}}}
\newcommand{\dphill}{\ensuremath{\Delta\phi_{\ell\ell}}}
\newcommand{\metrel}{\ensuremath{E_{\rm T,rel}^{\rm miss}}}
\newcommand{\pTll}{\ensuremath{p_{\rm T}^{\ell\ell}}}
\newcommand{\ptll}{\pTll}
\newcommand{\mT}{\ensuremath{m_{\rm T}}}

\newcommand{\hgg}{\ensuremath{H{\rightarrow\,}\gamma\gamma}}
\newcommand{\hww}{\ensuremath{H \rightarrow WW^{*}}}

\newcommand{\hwwlnln}{\ensuremath{H{\rightarrow\,}WW^{*}{\rightarrow\,}\ell\nu\ell\nu}}

\newcommand{\hzz}{\ensuremath{H \rightarrow ZZ^{*}}}

\newcommand{\hzzllll}{\ensuremath{H \rightarrow ZZ^{*}\rightarrow 4\ell}}

\newcommand{\Wjets}{$W$+jets}

\def\vec#1{{\mbox{$\boldsymbol{#1}$}}}

\def\thetast{\ensuremath{\theta^*}}
\def\costs{\ensuremath{|\cos \theta^*|}}
\newcommand{\mgg}{\ensuremath{m_{\gamma\gamma}}}

\usepackage{amssymb}
\usepackage{lineno}
\usepackage{chngpage}
\biboptions{numbers,square,comma,sort&compress}
\usepackage{hypernat}

\usepackage{preprintcover}  
\PreprintCoverPaperTitle{Evidence for the spin-0 nature of the Higgs boson using  ATLAS data}  
\PreprintIdNumber{CERN-PH-EP-2013-102}  
\PreprintCoverAbstract{Studies of the spin and parity quantum numbers of the Higgs boson are presented, based on
 proton--proton collision data collected by the ATLAS experiment at the LHC.
The Standard Model spin--parity \spinzero\ hypothesis is compared with alternative hypotheses 
using the Higgs boson decays \hgg ,  \hzzllll\ and \hwwlnln , as well as the combination of these channels. 
The analysed dataset corresponds to an integrated luminosity of 20.7~\ifb\ collected at a 
centre--of--mass energy of $\sqrt{s} = 8$ TeV. For the \hzzllll\  decay mode the dataset corresponding to 
an integrated luminosity of 4.6~\ifb\ collected at $\sqrt{s} = 7$ TeV is added. 
The data are compatible with the Standard Model $J^{P}= \spinzerop$ quantum numbers for the Higgs boson, whereas  
all alternative hypotheses studied in this letter, namely some specific $J^{P} =0^-,  ~1^+,~ 1^-,~ 2^{+}$ models, 
are excluded at confidence levels above 97.8\%. 
This exclusion holds independently of the assumptions on the coupling strengths 
to the Standard Model particles and in the case of the $J^{P} =2^+$ model, of the relative fractions of gluon--fusion and 
quark--antiquark production of the spin-2 particle.
The data thus provide evidence for the spin-0 nature of the Higgs boson, with positive parity being strongly preferred.}  
\PreprintJournalName{Physics Letters B}

\usepackage{hyperref}
 
\newcommand{\papertitle}{Evidence for the spin-0 nature of the Higgs boson using  ATLAS data}

\clearpage

\begin{document}

\title{\papertitle}

\author{ATLAS Collaboration}

\begin{frontmatter}
\begin{abstract}
Studies of the spin and parity quantum numbers of the Higgs boson are presented, based on
 proton--proton collision data collected by the ATLAS experiment at the LHC.
The Standard Model spin--parity \spinzero\ hypothesis is compared with alternative hypotheses 
using the Higgs boson decays \hgg ,  \hzzllll\ and \hwwlnln , as well as the combination of these channels. 
The analysed dataset corresponds to an integrated luminosity of 20.7~\ifb\ collected at a 
centre--of--mass energy of $\sqrt{s} = 8$ TeV. For the \hzzllll\  decay mode the dataset corresponding to 
an integrated luminosity of 4.6~\ifb\ collected at $\sqrt{s} = 7$ TeV is added. 
The data are compatible with the Standard Model $J^{P}= \spinzerop$ quantum numbers for the Higgs boson, whereas  
all alternative hypotheses studied in this letter, namely some specific $J^{P} =0^-,  ~1^+,~ 1^-,~ 2^{+}$ models, 
are excluded at confidence levels above 97.8\%. 
This exclusion holds independently of the assumptions on the coupling strengths 
to the Standard Model particles and in the case of the $J^{P} =2^+$ model, of the relative fractions of gluon--fusion and 
quark--antiquark production of the spin-2 particle.
The data thus provide evidence for the spin-0 nature of the Higgs boson, with positive parity being strongly preferred.
\end{abstract}
\end{frontmatter}

\hyphenation{ATLAS}

\section{Introduction} 
\label{sec:Introduction}
In 2012 the ATLAS and CMS Collaborations published the discovery of a new resonance~\cite{HiggsObservationATLAS, HiggsObservationCMS} in the search 
for the Standard Model (SM) Higgs boson $H$~\cite{Englert:1964et,Higgs:1964ia,Higgs:1964pj,Guralnik:1964eu,Higgs:1966ev,Kibble:1967sv}.
The present experimental challenge is to compare its properties with the SM predictions for the Higgs boson. 
In the SM, the Higgs boson is a spin-0 and CP-even particle ($J^{P}=0^+$). 
The Landau--Yang theorem forbids the direct decay of an on-shell spin-1 particle into a pair of 
photons~\cite{Landau,Yang}. The spin-1 hypothesis is therefore strongly disfavoured by the observation of the \hgg\ decay. 
The CMS Collaboration has published a spin--parity study~\cite{CMSzzSpin} based on the \hzz\ channel where the SM scalar hypothesis is favoured 
over the pseudoscalar hypothesis at a confidence level (CL) above 95\% .

In this Letter the \spinzero\ hypothesis of the SM is compared to several alternative hypotheses with
$J^{P} = 0^-, \; 1^+,\; 1^-,\; 2^{+}$. 
The measurements are based on the kinematic properties of the three final states \hgg, \hzzllll\ and \hwwlnln, where $\ell$ denotes an electron or a muon.
To improve the sensitivity to different spin--parity hypotheses, several final states are combined.
To test the \spinzerom\ spin--parity hypothesis, only the \hzz\ decay mode is used, while for the \spinonep\ and \spinonem\ hypotheses the 
\hzz\ and \hww\ channels are combined. For the \spintwop\ study, all three decay modes are combined.
 
The full dataset collected at $\sqrt{s}=8$ TeV, corresponding to an integrated luminosity of $20.7\;{\rm fb} ^{-1}$, is analysed for all three channels.  
For the \hzz\ decay mode, a dataset corresponding to an integrated luminosity of $4.6\;{\rm fb} ^{-1}$ collected at $\sqrt{s}=7$ TeV is also included. 

While for the SM Higgs boson the Lagrangian structure and its couplings are fully determined, the alternative hypotheses 
can be described by a wide variety of models, characterised by different structures and effective couplings. 
Several approaches to describe such signatures can be found in the literature~\cite{jhu1,jhu2,DeRuhula_lookAlikes,Choi03,deAquino:2011ix,Alwall:2011uj}.
In this Letter, the alternative model descriptions are based on Ref.~\cite{jhu1}, as described in Section~\ref{sec:SpinMC}. 
In Ref.~\cite{jhu1}, the production and decay of a generic boson with various $J^{P}$ quantum numbers are described 
by defining the most general amplitudes consistent with Lorentz invariance, angular-momentum conservation, Bose symmetry and 
the unbroken symmetry of the $SU(3) \times SU(2) \times U(1)$ gauge group. 
 
This Letter is published together with another one~\cite{ATLAScouplings} reporting the ATLAS measurements of the couplings 
of the Higgs boson derived from the observed signal production and decay rates. 
In that Letter the measurement of the mass of the Higgs boson, based on the invariant mass spectra in the \hgg\ and 
\hzzllll\ final states, is also reported.  
On the basis of that measurement, the observed final states are assumed to be produced in the decay of a single particle with a mass of 125.5~GeV~\cite{ATLAScouplings}. 
The definitions of the physics objects used in the analyses, the simulation of the different backgrounds and the main systematic uncertainties are  
described in Ref.~\cite{ATLAScouplings}. This Letter reports only aspects specific to the spin and parity analyses. 
The ATLAS collaboration has made public a collection of conference notes that document in detail the 
analyses reported in this Letter~\cite{conf_gamma, conf_4l, conf_ww}.

The outline of this Letter is as follows: Section~\ref{sec:SpinMC} describes the spin--parity models considered in all three channels and the signal 
Monte Carlo (MC) simulation samples used in the analyses. 
The statistical procedure used to test the different spin--parity hypotheses is presented in Section~\ref{sec:Statistics}.
Sections~\ref{sec:hgg},~\ref{sec:hzz} and ~\ref{sec:hww} provide brief descriptions of the spin--parity analyses in the \hgg, \hzz\ and \hww\ decay modes. 
Finally, in  Section~\ref{sec:Results}, the combined results in terms of compatibility with several spin--parity hypotheses are presented.

\section{Signal modelling and Monte Carlo samples}
\label{sec:SpinMC} 
The interactions of spin-0, 1 and 2 resonances with Standard Model particles 
are described in Ref.~\cite{jhu1} by Eqs.~2, 4 and 5 for bosons and by Eqs.~8, 9 and 10 for fermions.
The choices of the boson and fermion couplings for the specific spin and parity 
models used in this analysis are presented in Table 1 of Ref.~\cite{jhu1}. 

The implications of these choices are briefly summarised in the following.
The quark--antiquark ($q\bar{q}$) annihilation production process is not considered in the case of \spinzeromm , since 
its contribution is negligible compared to gluon fusion ($gg$).
For the $J^{P} = 1^{+}$ and $1^{-}$ cases, only the quark--antiquark annihilation production process is considered, 
since the Landau--Yang theorem also forbids the production of a spin-1 particle through the fusion of two on-shell gluons.
Given the large number of possible spin-2 models, a specific one, denoted by $2^{+}_{m}$ from Table 1 of Ref.~\cite{jhu1}, was chosen.
This choice corresponds to a graviton-inspired tensor with minimal couplings to SM particles. 
In the $2^{+}_{m}$ boson rest frame, its polarisation states projected onto the parton collision axis can take only the values of $\pm 2$  for the gluon-fusion 
process and $\pm 1$ for the quark--antiquark annihilation process. For the spin-2 model, only these two production mechanisms are considered.
The production of the  $2^{+}_{m}$ boson is dominated by the gluon-fusion process with a contribution, at leading order in quantum chromodynamics (QCD), of about 4\% from
quark--antiquark annihilation~\cite{deAquino:2011ix,Alwall:2011uj}.
This proportion could be significantly modified by higher-order QCD corrections.
Since the experimental observables are sensitive to different polarisations, the studies were performed for several production admixtures 
by normalising the samples produced with the two different production processes in order to obtain samples of events corresponding to 
fractions, $f_{q\bar{q}}$, of $q\bar{q}$ annihilation ranging from $0\%$ to $100\%$ in steps of $25\%$. In the following, this model is referred to as \spintwo .

The production and decay of the SM Higgs boson via the dominant gluon-fusion process is simulated 
using either the JHU Monte Carlo generator~\cite{jhu1} for the \hzz\ process or   
the POWHEG~\cite{powheg} Monte Carlo generator for the \hgg\ and \hww\ processes, each interfaced to PYTHIA8~\cite{pythia8} for parton 
showering and hadronisation. 
The production and decay of the $J^P=0^{-}, 1^{+}, 1^{-}$ and $2^+$ resonances are modelled using the
JHU generator, interfaced to PYTHIA8 for parton showering and hadronisation. 

The transverse momentum (\pt) distributions for the gluon-fusion signals produced with the JHU generator, which is leading-order in QCD, are weighted to reproduce 
the POWHEG+PYTHIA8 spectrum.
The latter was tuned to reproduce the re-summed calculation of the HqT program~\cite{deFlorian:2011xf}.
It was checked that the distributions of all kinematic variables used for the spin--parity determination
are compatible between the two MC generators after the re-weighting is applied. 
For the production process via $q\bar{q}$ annihilation, no re-weighting is applied.

The much smaller contributions from other production processes, namely vector-boson fusion and associated production, 
are also considered. For the \hgg\ channel, they are included in the analysis and simulated as described in Ref.~\cite{ATLAScouplings}. 
For the  \hzz\ channel, they are ignored because they do not affect the kinematic distributions used in the spin analysis. 
For the \hww\ analysis, where only the $e \mu$ final state with no additional jet activity is considered, as described in Section~\ref{sec:hww}, 
they contribute at a negligible level and are therefore ignored.

For the background processes, the simulated samples are the same as those used in the coupling analyses. 
A detailed list of the MC generators and samples is given in Ref.~\cite{ATLAScouplings}.

All MC samples are passed through a full simulation of the ATLAS detector~\cite{atlasSim} based on GEANT4~\cite{Geant4}. 
The simulation incorporates a model of the event pile-up conditions in the data, including the effects of multiple proton--proton collisions in in-time and nearby bunch crossings.

\section{Statistical method} 
\label{sec:Statistics}
The analyses described in this Letter rely on discriminant observables chosen to be sensitive to the spin and parity of the signal 
while preserving the discrimination against the various backgrounds, as described 
in Sections~\ref{sec:hgg},  ~\ref{sec:hzz} and  ~\ref{sec:hww} for the three final states. 
A likelihood function ${\cal L}(\spin, \mu, \vec{\theta})$ that depends on the spin--parity assumption of the signal 
is constructed  as a product of conditional probabilities over binned distributions
 of the discriminant observables in each channel:

\begin{equation}
\begin{split}
& \mathcal{L}(\spin, \mu, \vec{\theta}) =  \displaystyle  \prod_{j}^{N_{\mathrm{chann.}}} \prod_{i}^{N_{\mathrm{bins}}} \\
& P\big(N_{i,j} ~| ~\mu_{j} \cdot S^{(\spin)}_{i, j}(\vec{\theta}) ~+ B_{i,j}(\vec{\theta}) \big)  \times  \mathcal{A}_{j}(\vec{\theta}) \;,
\end{split}
\label{eq:likelihood}
\end{equation}

\noindent where $\mu_{j}$ represents the nuisance parameter associated with the signal rate in each channel $j$. 
The symbol $\vec{\theta}$ represents all other nuisance parameters. 
The likelihood function is therefore a product of Poisson distributions $P$ corresponding 
to the observation of $N_{i,j}$ events in each bin $i$ of the discriminant observable(s),\footnote{As explained in the following sections, 
the sensitivity for spin--parity separation is improved by a  simultaneous fit to two discriminants in the \hgg\ and \hww\
decay modes, while in the \hzz\ channel only one discriminant is used.}
given the expectations for the signal, $S_{i,j}^{(\spin)}(\vec{\theta})$, and for the background, $B_{i,j}(\vec{\theta})$. 
Some of the nuisance parameters are constrained by auxiliary measurements through the functions $\mathcal{A}_{j}(\vec{\theta})$.

While for the SM Higgs boson the couplings to the SM particles are predicted, they are not known \textit{a priori} for the alternative hypotheses, 
defined as \spinalt . 
In order to be insensitive to such assumptions, the numbers of signal events in each channel and for each tested hypothesis are treated as an independent 
nuisance parameters in the likelihood.

The test statistic $q$ used to distinguish between the two signal spin--parity hypotheses is based on a ratio of likelihoods:

\begin{equation}
q = \log \frac{\mathcal{L}(\spinzero ,  \hat{\hat{\mu}}_{\spinzerop}, \hat{\hat{\theta}}_{\spinzerop}) } {\mathcal{L}(\spinalt ,  \hat{\hat{\mu}}_{\spinalt}, \hat{\hat{\theta}}_{\spinalt} )}\;,
\label{eq:lambda}
\end{equation}

\noindent where $\mathcal{L}(\spin,  \hat{\hat{\mu}}_{\spin}, \hat{\hat{\theta}}_{\spin})$ is the maximum likelihood estimator, evaluated under 
either the \spinzerop\ or the \spinalt\ spin--parity hypothesis. The $\hat{\hat{\mu}}_{\spin}, ~\hat{\hat{\theta}}_{\spin}$ represent the values of the 
signal strength and nuisance parameters fitted to the data under each $J^{P}$ hypothesis.
The distributions of the test statistics for each of the two hypotheses are obtained using ensemble tests (Monte Carlo pseudo-experiments). 
The generation of the pseudo-experiments uses the numbers of signal and background events in each channel obtained 
from maximum likelihood fits to data. In the fits of each pseudo-experiment, these and all other nuisance parameters are profiled, i.e. fitted to 
the value that maximises the likelihood for each value of the parameter of interest. 
When generating the distributions of the test statistics for a given spin--parity hypothesis, the signal strength $\mu$ is fixed to the 
value obtained in the fit to the data under the same spin--parity assumption.
The distributions of $q$ are used to determine the corresponding $p_0$-values $p_0(\spinzerop)$ and $p_0(\spinalt)$. 
For a tested hypothesis \spinalt ,   the observed (expected) $p_0$-values are obtained by integrating the corresponding test-statistic 
distributions above the observed value of $q$ (above the median of the \spinzero\ $q$ distribution).
When the measured data are in agreement with the tested hypothesis, the observed value of $q$ is expected to be close to 
the median, corresponding to a $p_0$-value around $50\%$. 
Very small values of the integral of the \spinalt\ distribution, corresponding to large values of $q$, are interpreted as the data 
being in disagreement with the tested hypothesis in favour of the SM hypothesis. 
An example of such distributions is shown in Section~\ref{sec:Results} for the \spinzerop\ and \spinzerom\ hypotheses. 

The exclusion of the alternative \spinalt\ hypothesis in favour of the Standard Model \spinzerop\ hypothesis is evaluated in terms 
of the corresponding $\CLs(\spinalt)$, defined as:

\begin{equation}
\CLs(\spinalt) = \frac{p_0 (\spinalt)}{ 1- p_0 (\spinzerop)}\;.
\label{eq:cls}
\end{equation}

\section{\hgg\  Analysis} 
\label{sec:hgg}
The \hgg\ decay mode is sensitive to the spin of the Higgs
boson through the measurement of the polar angular distribution of the
photons in the resonance rest frame. 
For this channel, the SM spin hypothesis is compared only 
to the \spintwo\ hypothesis. 
Spin information can be
extracted from the distribution of the absolute value of the cosine of
the polar angle \thetast\ of the photons with respect to the $z$-axis
of the Collins--Soper frame~\cite{CollinsSoper}:

\begin{equation}
  \costs\ = \frac{|\sinh(\Delta\eta^{\gamma\gamma})|}{\sqrt{1+(\pT^{\gamma\gamma}/\mgg)^2}}\frac{2\pT^{\gamma1}\pT^{\gamma2}}{\mgg^2}\; ,
  \label{eq:hgg_costs}
\end{equation}

\noindent where $\mgg$ and $\pT^{\gamma\gamma}$ are the invariant mass and the transverse momentum of the photon pair, $\Delta\eta^{\gamma\gamma}$ 
is the separation in pseudo-rapidity of the two photons, and $\pT^{\gamma1},\pT^{\gamma2}$ are the transverse momenta of the photons.

This channel has a large background, dominated by 
non-resonant diphoton production, whose distribution in \costs\
is intermediate between those expected for $J^{P} = 0^+$ and
$J^{P} = 2^+$ states produced in gluon fusion. 
Two observables, \costs\ and \mgg, are used in the fit to data: 
\mgg\ provides better separation power between the signal and the background, 
and \costs\ is sensitive to the spin.

The selected events contain two isolated photon candidates, 
as described in Ref.~\cite{ATLAScouplings}, but with 
the important difference that the kinematic requirements on the transverse momenta of the photons are proportional
to \mgg . This choice reduces the correlation between \mgg\ and \costs\ for the background to a negligible level. 
The selection requirements are set to
$\pT^{\gamma1}>0.35~\mgg$ and $\pT^{\gamma2}>0.25~\mgg$. 
The fitted mass range is chosen to be $105\gev<\mgg<160\gev$.

The intrinsic width of the resonance is assumed to be
negligible compared to the detector resolution for both spin hypotheses.
For this reason, the same probability density function (pdf) is used to 
model the reconstructed mass spectra of both signal hypotheses, independent of the value of \costs. The chosen function is
the sum of a Crystal Ball~\cite{crystalball2} component, accounting for about 95\% of the
signal events, and a wider Gaussian component to model outlying
events, as described in Ref.~\cite{ATLAScouplings}. 

The \costs\ distributions of the signal, for either spin state, 
are obtained from simulated samples. The signal yields per \costs\ bin for a spin-0 particle are corrected for
interference effects with the non-resonant diphoton background ${gg
\to \gamma\gamma}$~\cite{Dixon:2003yb}. 
The size of the correction is non-negligible only at high values of \costs\ and its value is taken as the systematic uncertainty on this effect.
No interference between the spin-2
particle and the diphoton continuum background is assumed, since there are
no theoretical models that describe it. 

For the spin-2 state, the full size of the correction to the generated \pT\ spectrum of the diphoton system, described in Section 2, 
is taken as a systematic uncertainty.

The background distributions are derived directly from the observed data,
using the two mass sidebands $105\gev<\mgg<122\gev$ and $130\gev<\mgg<160\gev$, 
where the signal contribution is negligible. The background shape as a function of \mgg\ is modeled by a
fifth-order polynomial with coefficients fitted to the data.
The background shape as a function of \costs\ is taken from the two mass sidebands, 
since the remaining correlation between the two observables is small.
The statistical uncertainties affecting the determination of the \costs\ distribution from 
the sidebands are propagated into the signal region (SR), $122\gev<\mgg<130\gev$,
independently for each \costs\ bin. 
Detailed studies of the data in the sidebands show that
possible residual correlations between \mgg\ and \costs\ are not significant compared
to the statistical uncertainties. A study of the background, based on a large sample of simulated events using the SHERPA 
generator~\cite{Sherpa}, indicates the presence of a residual correlation at the level of 0.6\% for $\costs<0.8$ and 2\% elsewhere.
These values are treated as the systematic uncertainties due to possible correlations between 
\mgg\ and \costs .

The fit to data is carried out simultaneously in the signal region and the two sideband regions.
In the signal region, the likelihood is a function of the two discriminant variables $\mgg$ and  $\costs$, 
while in the sidebands only \mgg\ is considered.

\begin{figure}[!htbp]
  \centering
  \includegraphics[width=1.0\columnwidth]{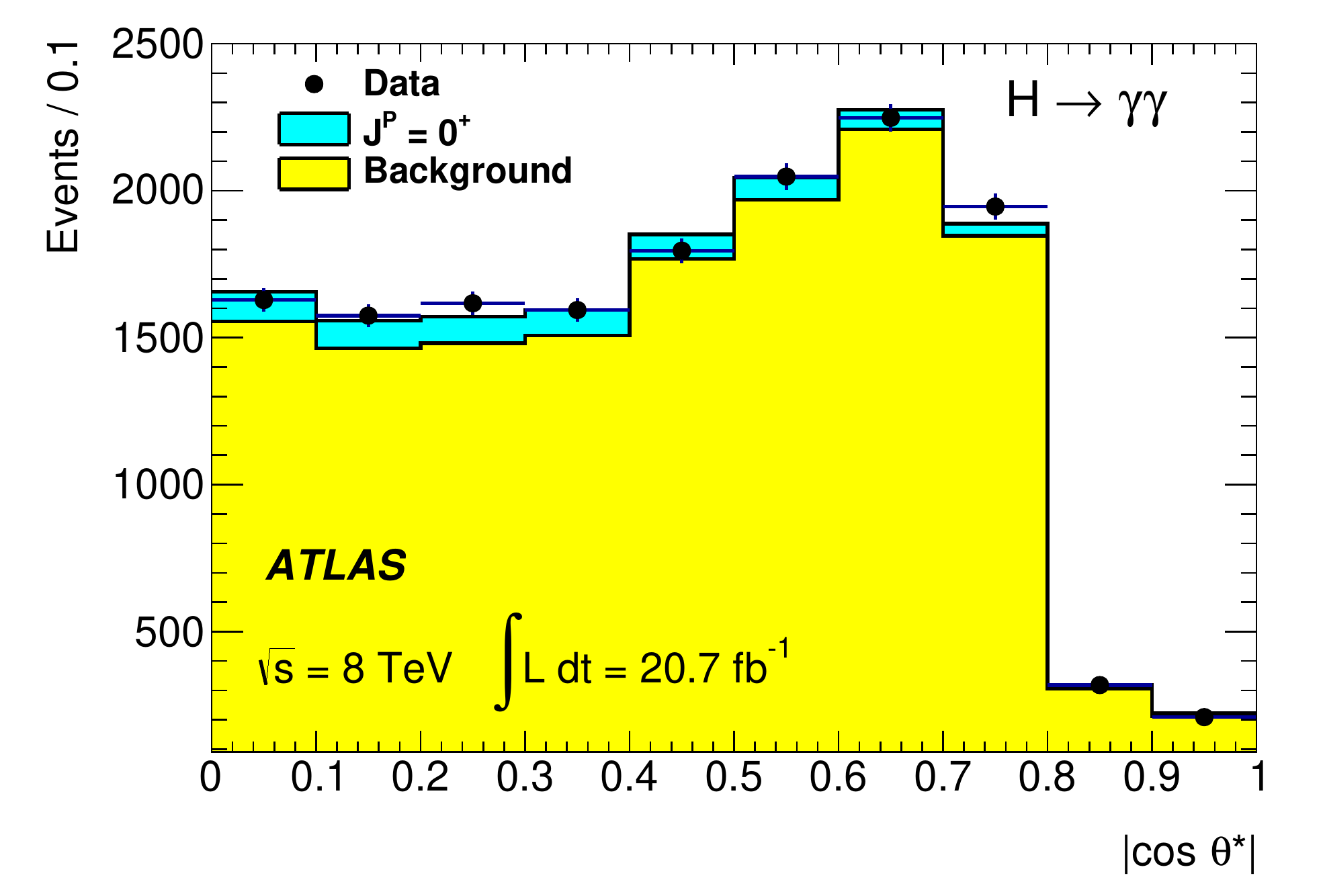}
  \caption{
    Distribution of \costs\ for events in the signal region defined by $122\gev<\mgg<130\gev$.  The data (dots) are 
    overlaid with the projection of the signal
    (blue/dark band) and background (yellow/light histogram) components obtained from the inclusive fit of the data under the spin-0 hypothesis.
  }
  \label{fig:Hgg_spin_costheta_SR_spin0}
\end{figure}

The number of data events selected in the signal region is 14977, compared with 
a background estimate of about 14300 events and an expected SM Higgs boson signal of about 370 events.
Figure~\ref{fig:Hgg_spin_costheta_SR_spin0} displays the data distribution for \costs\ in the signal region, 
overlaid with the signal and background components, fitted under the \spinzero\ hypothesis.

\begin{figure}[!htb]
  \centering
  \subfigure[]{\includegraphics[width=1.0\columnwidth]{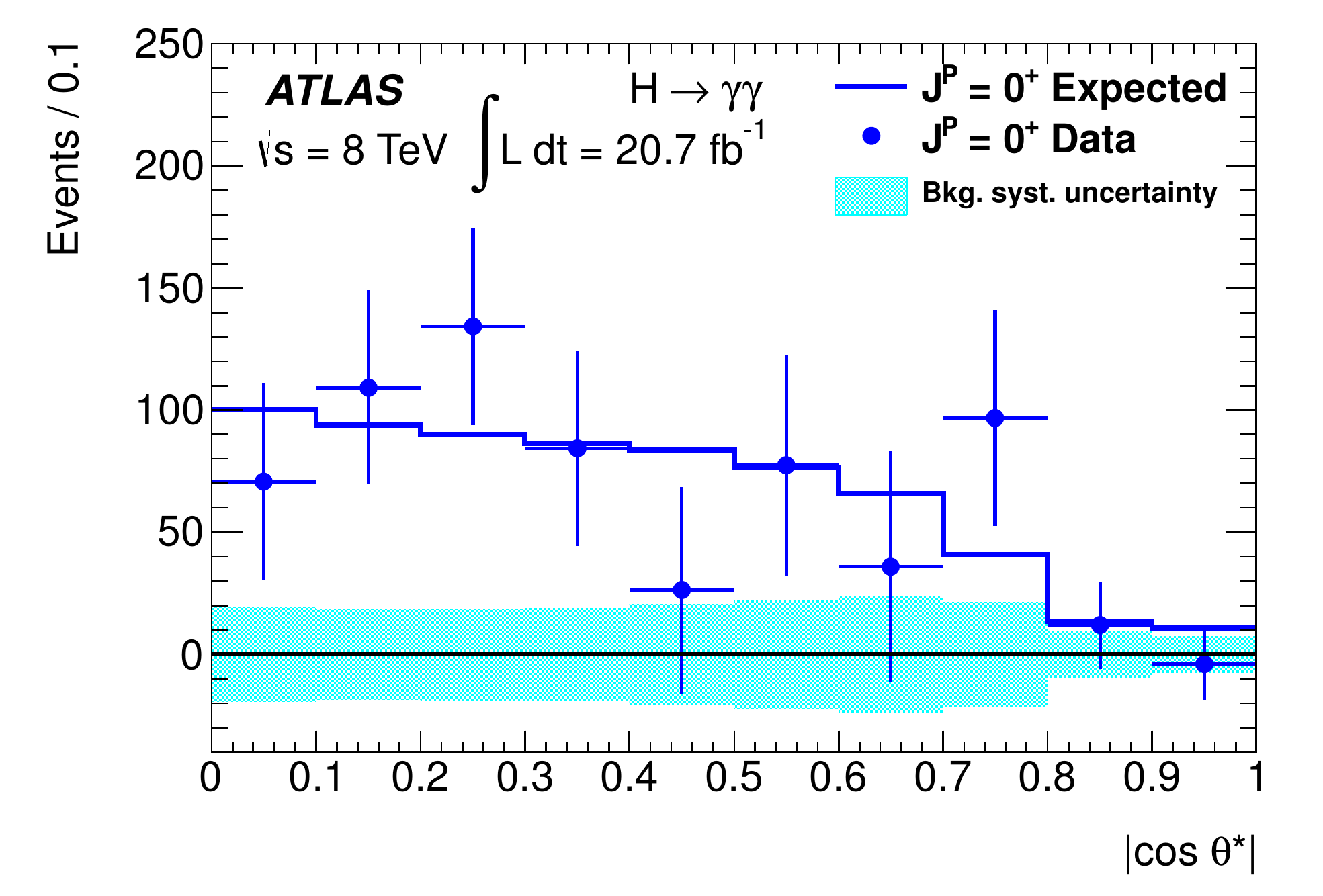}}
  \subfigure[]{\includegraphics[width=1.0\columnwidth]{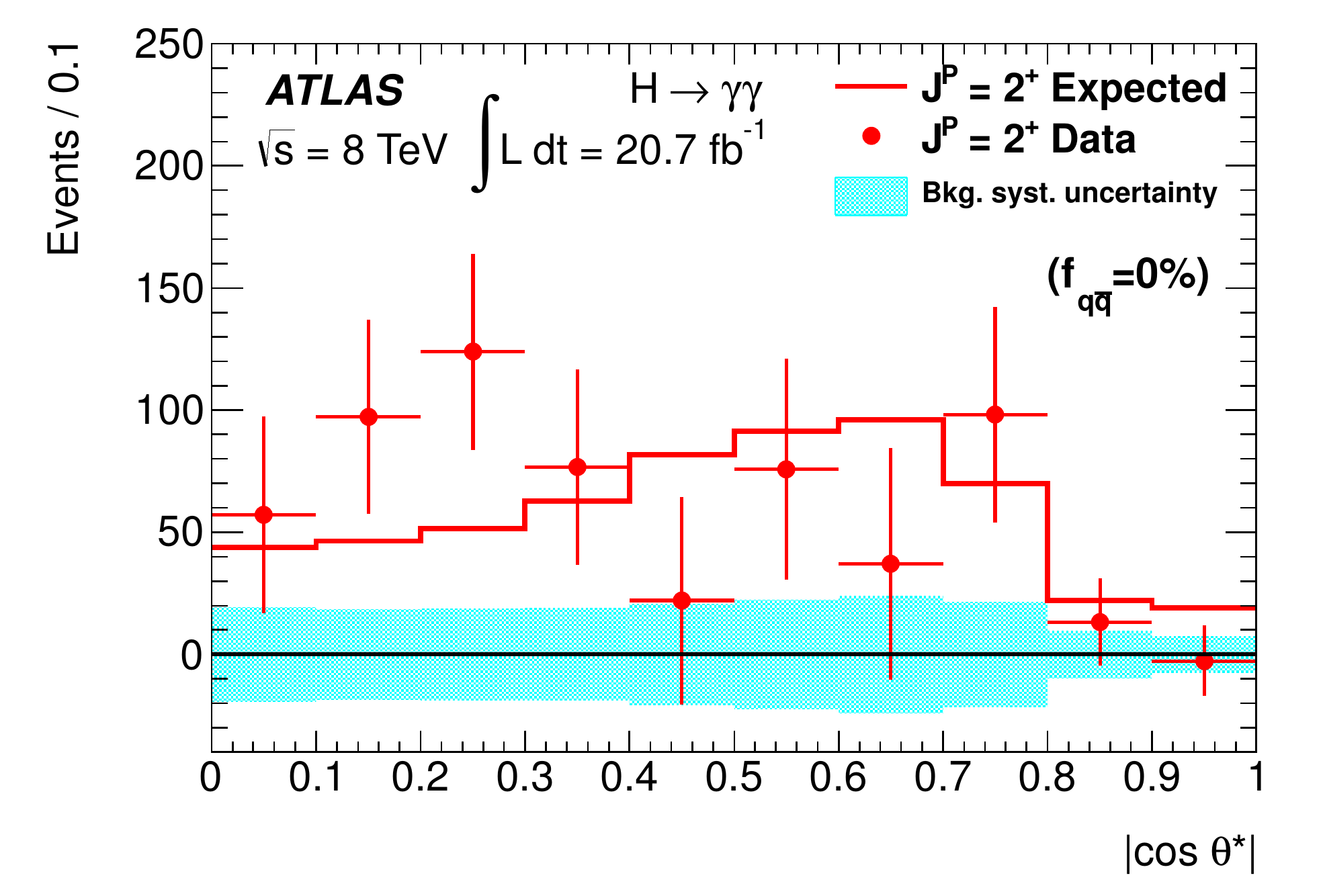}}  
  \caption{
    Distributions of background-subtracted data in the signal region
    as a function of $\costs$. The expected distributions for (a) spin-0 and (b) spin-2 signals produced by gluon fusion,
    normalised to the fitted number of signal events, are overlaid as solid lines.
    The cyan/grey bands around the
    horizontal lines at zero show the systematic uncertainties on the
    background modelling before the fits, which
    include the statistical uncertainties on the data sidebands. 
  }
  \label{fig:Hgg_spin_costheta_SR_bkgsubtracted}
\end{figure}

The likelihood function is fitted to data for both the spin-0 and spin-2 hypotheses with the signal and 
background normalisations treated as nuisance parameters.
Figure~\ref{fig:Hgg_spin_costheta_SR_bkgsubtracted} shows the \costs\ distributions in the signal region, 
obtained after subtracting the estimated background, and compared with the expected distributions
for spin-0 and spin-2 signals. 
The data points differ slightly between the two spin hypotheses,
because the fitted background depends on 
the profiling of the nuisance parameters associated with the 
bin-by-bin systematic uncertainties.

\section{\hzzllll\ Analysis} 
\label{sec:hzz}
The \hzzllll\ channel, where $\ell\ = e$ or $\mu$, benefits from the presence of several observables dependent 
on spin and parity thanks to the full reconstruction of the four-lepton final state.
The kinematic observables are the reconstructed masses of the two $Z$ boson candidates and the 
five production and decay angles described in the following. 
The $Z$ boson candidates are denoted hereafter as $Z_1$ and $Z_2$,
where the index $1$ refers to the lepton pair with the invariant mass closer to the PDG value~\cite{pdg} of the $Z$ boson mass. 
Their respective masses are defined as $m_{12}$ and $m_{34}$.
The full definition  of the production and decay angles  as well as the description of their variation
for different spin and parity values can be found in Ref.~\cite{conf_4l}.
Here only a brief summary is given: $\theta_1$ ($\theta_2$) is the angle between the negatively charged final-state 
lepton in the $Z_1$ ($Z_2$) rest frame and the direction of flight of the $Z_1$ ($Z_2$) boson in the four-leptons rest frame. 
$\Phi$ is the angle between the decay planes defined by the two lepton pairs coming from the $Z$ decays
 in the four-lepton rest frame. $\Phi_1$ is the angle between the decay plane of the leading lepton pair and a 
plane defined by the momentum of the $Z_1$ in the four-lepton rest frame and the direction of the
beam axis. $\theta^*$ is the production angle of the $Z_1$ defined in the four-lepton rest frame.

The lepton identification criteria and the analysis requirements follow the inclusive event selection described 
in Ref.~\cite{ATLAScouplings}. To increase the 
sensitivity to the Higgs boson signal the final states are classified depending on the flavours of the lepton pairs.
The events used to reconstruct the variables sensitive to the spin and parity of the resonance are selected in 
the region of reconstructed four-lepton invariant mass $115\;{\rm GeV}<m_{4 \ell}<130\;{\rm GeV}$, 
defined as the signal mass window.

\begin{figure}[!htbp]
\centering
\subfigure[]{\includegraphics[width=1.0\columnwidth]{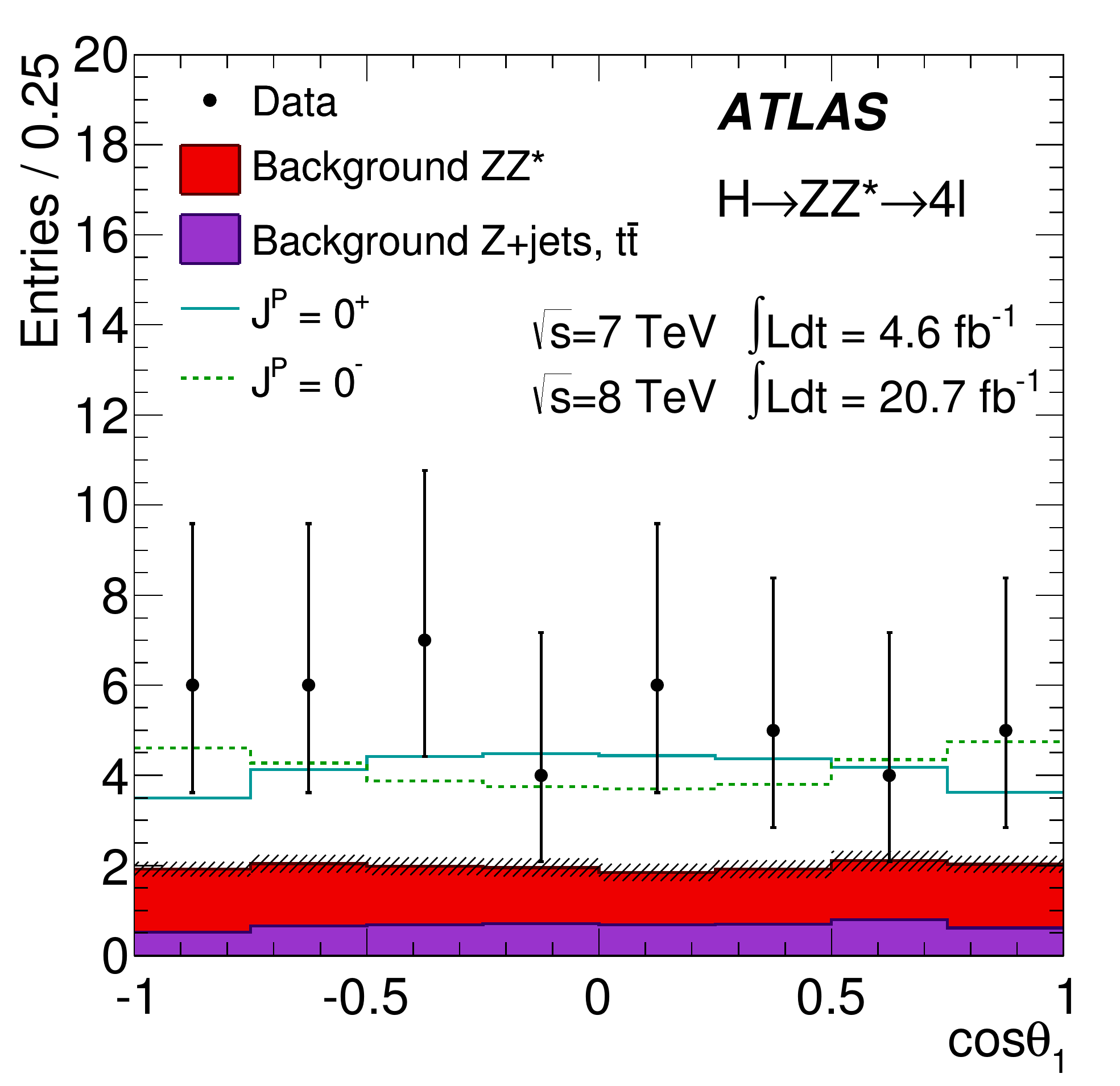}}
\subfigure[]{\includegraphics[width=1.0\columnwidth]{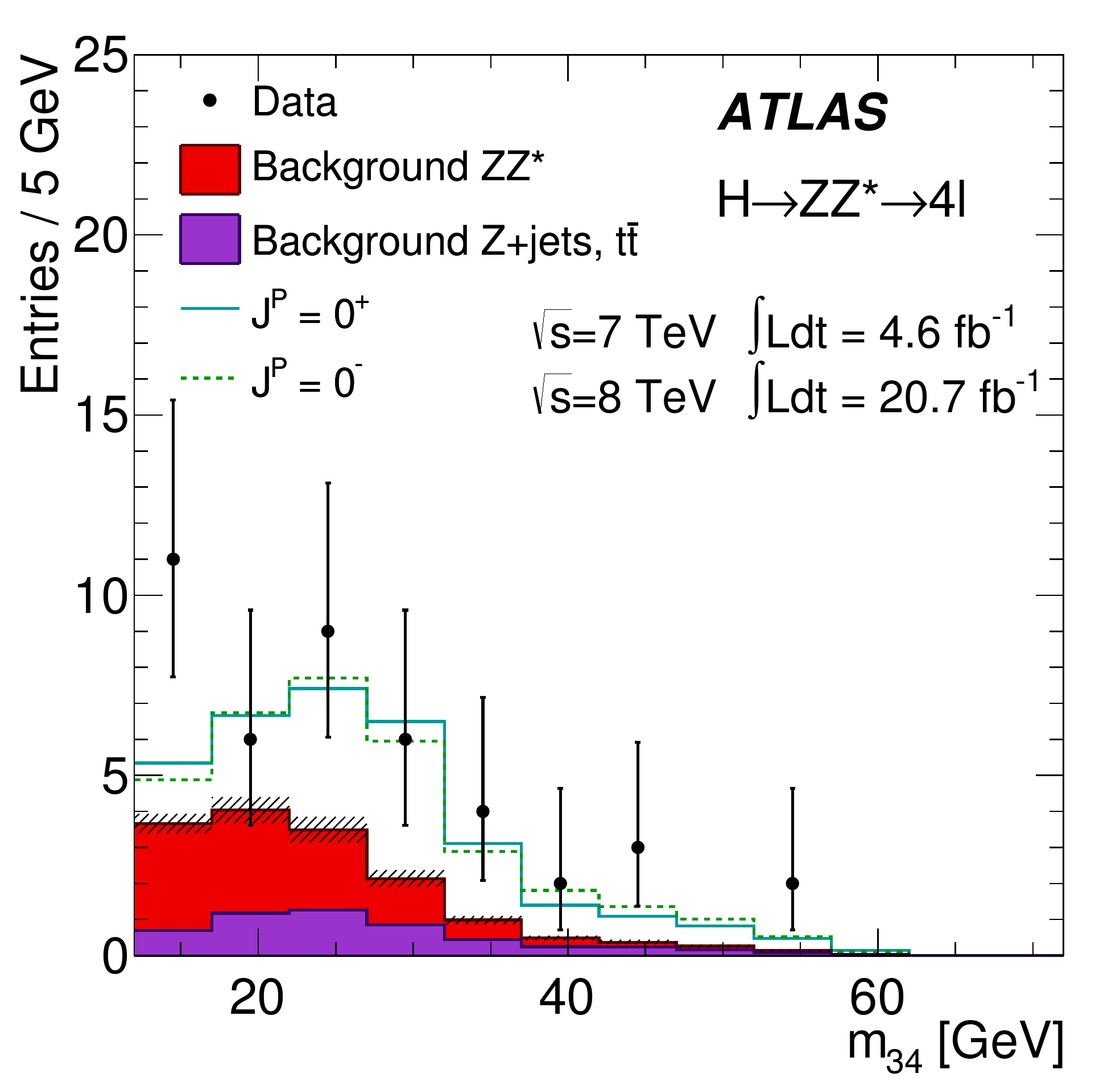}}
\caption{Distributions of (a) $\cos(\theta_1)$ and  (b) $m_{34}$
for events passing the full selection in the signal  mass window  $115\;{\rm GeV}<m_{4 \ell}<130\;{\rm GeV}$
for the combined $\sqrt{s}=7\;{\rm TeV}$ and  $\sqrt{s}=8\;{\rm TeV}$ datasets. 
The expected contributions from the  $J^{P}=0^+$ (solid line) and  $J^{P}=0^-$ (dashed line)
signal hypotheses, and the irreducible $ZZ^*$ background are shown, together with the measured contribution from reducible non-$ZZ^{*}$
backgrounds. The hatched areas represent the uncertainty on the background yields from statistical, experimental, and theoretical sources.}
\label{fig:ZZVars}
\end{figure}

After the analysis requirements 43 candidate events are selected in data in the signal mass window, compared with an expected background of about 16 events, 
dominated by the continuum $Z Z^{*}$ process, and about 18 signal events for a SM Higgs boson with a mass of $125.5 ~{\rm GeV}$.
The irreducible $Z Z^{*}$ background is estimated from Monte Carlo simulation, normalised to NLO calculations, while
the reducible $t\bar{t}$, $Zb\bar{b}$ and \Zjets\ backgrounds are estimated from corresponding control regions in data, as described in Ref.~\cite{ATLAScouplings}.
Figure~\ref{fig:ZZVars} shows the $\cos(\theta_1)$ and  $m_{34}$  distributions for events 
passing the full selection in the signal  mass window.

In order to distinguish between pairs of spin and parity states, the reconstructed observables  
described above, namely the five angles and the two invariant masses, 
are combined using a multivariate discriminant based on a boosted decision tree (BDT)~\cite{tmva2}.
The BDT is trained on simulated signal events after full reconstruction and event selection.
Dedicated discriminants are defined for the separation between the Standard Model \spinzero\ hypothesis and each of the considered alternative models, 
$J^{P} =0^-, \; 1^+,\; 1^-,\; 2^{+}$. 
In the case of the spin-2 hypothesis, the studies are performed as a function of the $q\bar{q}$ production fraction, \fqq .

\begin{figure}[!htbp]
\centering
\subfigure[]{\includegraphics[width=1.0\columnwidth]{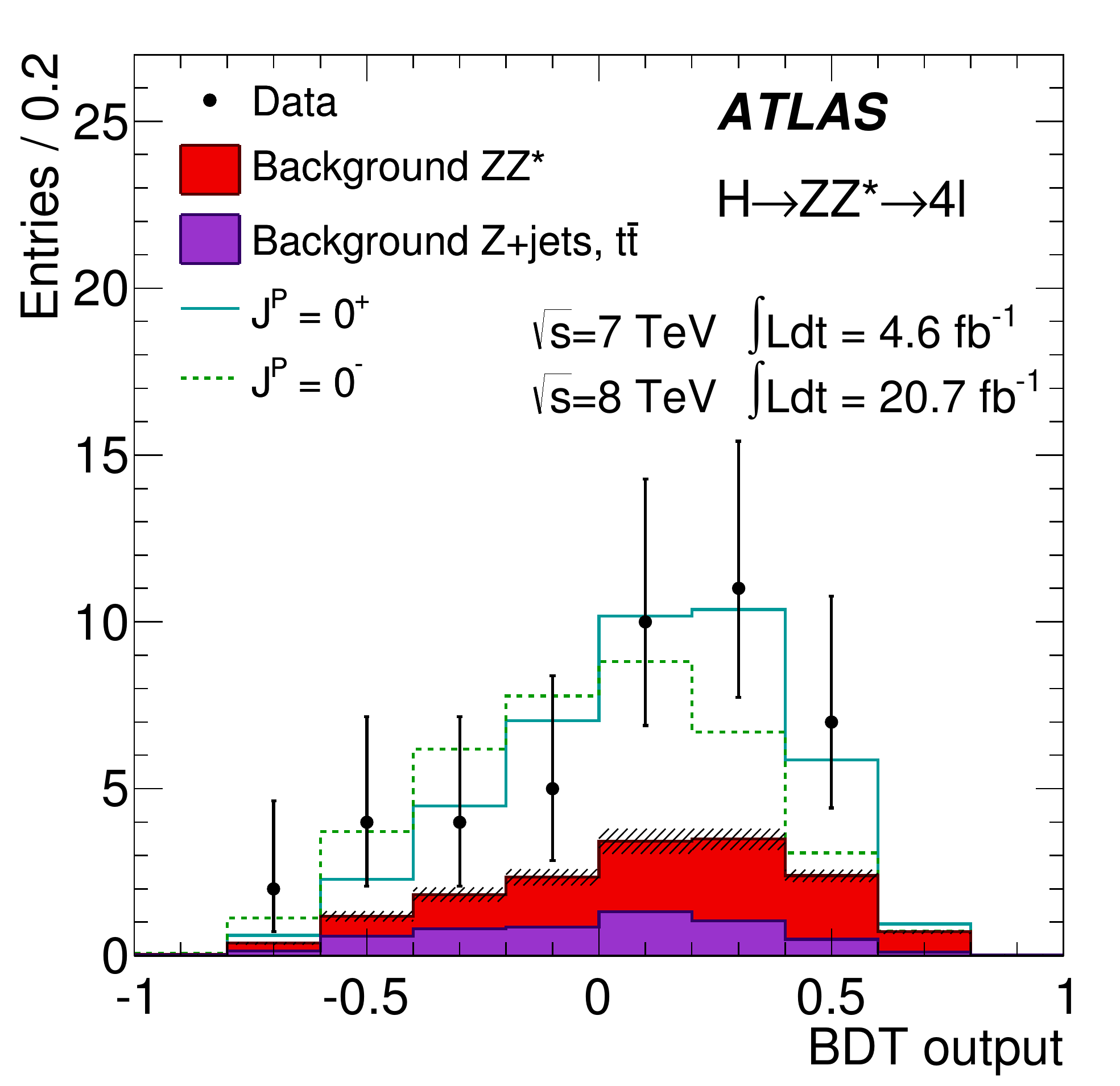}}
\subfigure[]{\includegraphics[width=1.0\columnwidth]{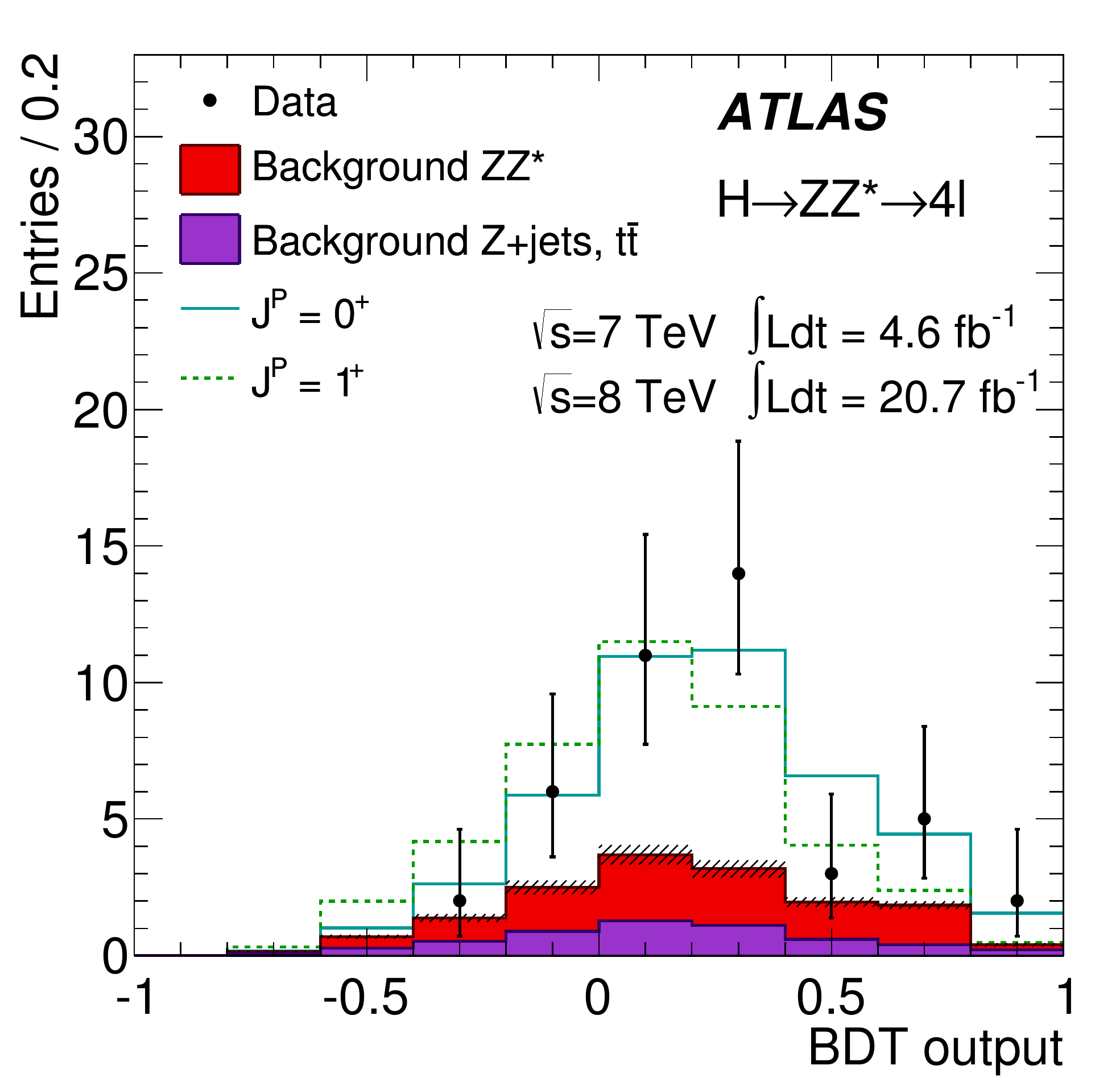}}
\caption{Distributions of the BDT output for data (points with error bars) and expectations based on MC simulation (histograms).
The distribution of each discriminant is shown for a pair of spin and parity hypotheses for the signal:
$J^{P}=0^+$ (solid line) and  $J^{P}=0^{-}$ (dashed line) in (a), $J^{P}=0^+$ (solid line) and  $J^{P}=1^+$ (dashed line)
in (b). The signal contribution for each of the two hypotheses is scaled using the profiled value of the signal strength.
The hatched areas represent the uncertainty on the background yields from statistical, experimental, and theoretical sources.}
\label{fig:ZZDiscr}
\end{figure}

The response of the BDT classifiers is evaluated separately for each pair of signal hypotheses, including the expected backgrounds from other SM processes.
In addition, to improve the overall sensitivity,
the BDT responses are evaluated separately for two $m_{4 \ell}$ regions with high and low signal-over-background ratio (S/B): low (115--121 GeV
and 127--130 GeV) and high (121--127 GeV).

Systematic uncertainties on the shapes of the BDT output and on the normalisations of the high and low S/B mass regions are considered. 
These are due to uncertainties on the lepton identification efficiencies, the lepton energy scale and its resolution.
A systematic uncertainty of $\pm 10\%$ on the normalisation of the high and low S/B mass regions is applied to take into account the experimental 
uncertainty on the mass of the Higgs boson.
The systematic uncertainties on the overall background yields and on the integrated luminosity are treated as described in Ref.~\cite{ATLAScouplings}.
Figure~\ref{fig:ZZDiscr} shows the BDT discriminant distributions for the $J^{P}=0^+$ versus $J^{P}=0^-$ 
and the $J^{P}=0^+$ versus $J^{P}=1^+$ hypotheses. 
The distribution of the BDT output is used as a discriminant observable in the likelihood defined in Section~\ref{sec:Statistics}.

In addition to the BDT analysis an alternative approach based on the differential 
decay rate with respect to the angles and the masses, $m_{12}$ and $m_{34}$, was also studied. These variables, corrected for detector acceptance and analysis 
selection effects, are used to construct a matrix-element-based discriminant.
This alternative analysis yields results compatible with those obtained with the 
BDT, as described in detail in Ref.~\cite{conf_4l}.

\section{\hwwlnln\ Analysis}
\label{sec:hww}
The analysis of the spin and parity in the \hwwlnln\ channel is restricted 
to events containing two leptons of different flavour (one electron and one muon) 
and no observed jets with $\pt >25$~\GeV\ within $|\eta| < 2.5$ or with $\pt >30$~\GeV\  within ${2.5 < |\eta| < 4.5}$.  
The leading lepton is required to have $\pT>25$~\GeV\ and the sub-leading lepton $\pT>15$~\GeV. 
At least one of the two selected leptons is required to match a lepton that triggered the recording of the event.

The major sources of background after the dilepton selection are: $Z/\gamma^{*}$+jets, diboson ($WW$, $WZ/ \gamma^*$, $ZZ/ \gamma^{*}$), top-quark ($t\bar{t}$ and single
top) production, and $W$ bosons produced in association with hadronic jets where a jet is misidentified as a lepton.
The requirement of two high-$\pT$ isolated leptons significantly reduces the
background contributions from fake leptons. Multi-jet and $Z/\gamma^*$ events are suppressed 
by requiring relative missing transverse momentum\footnote{$\metrel \equiv \met \cdot \textrm{sin}\,\Delta\phi$, 
where $\Delta\phi$ is the azimuthal separation between the missing transverse momentum and the nearest 
reconstructed object (lepton or jet with $\pT > 25$~\GeV) or ${\pi}/{2}$, whichever is smaller. 
The missing transverse energy \met\ is defined as the modulus of the missing transverse momentum.} 
$\metrel$ above 20~\GeV.

Further lepton topological requirements are applied to optimise the
sensitivity for the separation of different spin hypotheses, namely requirements on the dilepton invariant mass $\mll < 80$~\GeV, 
the transverse momentum of the dilepton system $\ptll > 20$~\GeV\ and 
the azimuthal angular difference between leptons $\dphill < 2.8$~
radians. This selection, which significantly reduces the $WW$ continuum and $Z/\gamma^*$ backgrounds, defines the signal region (SR). 
 
The contributions from $WW$, top-quark and \Zjets\ processes  predicted by MC simulation are normalised to observed rates
in control regions (CRs) dominated by the relevant background sources. The \Zjets\ CR is defined by inverting the
\dphill\ requirement and removing the \ptll\ one. The \Zjets\ normalisation factor of 0.92 
with a total uncertainty of $\pm 8 \%$ is derived from this control region and applied to the simulated sample.  
The $WW$ CR is defined using the same selection as for the SR except that the \dphill\ requirement is removed and the \mll\ requirement
is inverted. The resulting $WW$ normalisation factor applied to the MC prediction is 1.08 with a total uncertainty of $\pm 10\%$. 
The top-quark background is estimated as described in Ref.~\cite{ATLAScouplings}. The ratio of the resulting prediction to the 
one from simulation alone is 1.07 with a total uncertainty of $\pm 14\%$. 
The \Wjets\ background is estimated entirely from data. The shapes and normalisations of non-$WW$ diboson backgrounds are estimated 
using simulation and cross-checked in a validation region~\cite{ATLAScouplings}. 
The correlations introduced among the different background sources by the presence of other processes in the 
control regions are fully included in the statistical procedure to test the compatibility between data and the two spin
hypotheses, as described in Section~\ref{sec:Statistics}.

After the selection, the data SR contains 3615 events, with 170 events expected from the SM Higgs boson signal and 
about 3300 events from background processes, after their normalisation to data in the CRs.

\begin{figure}[!htbp]
\centering
\subfigure[]{\includegraphics[width=1.0\columnwidth]{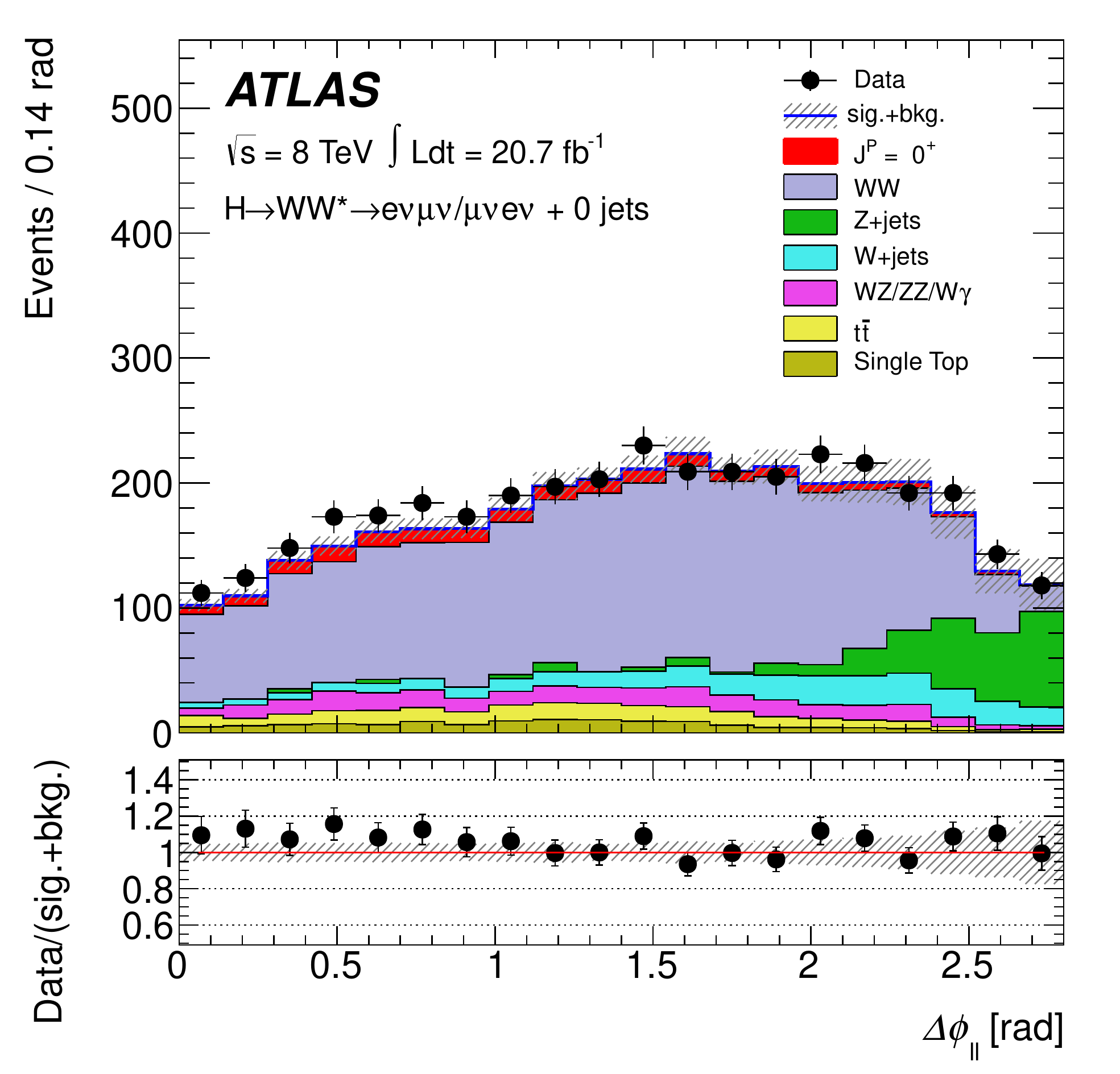}}
\subfigure[]{\includegraphics[width=1.0\columnwidth]{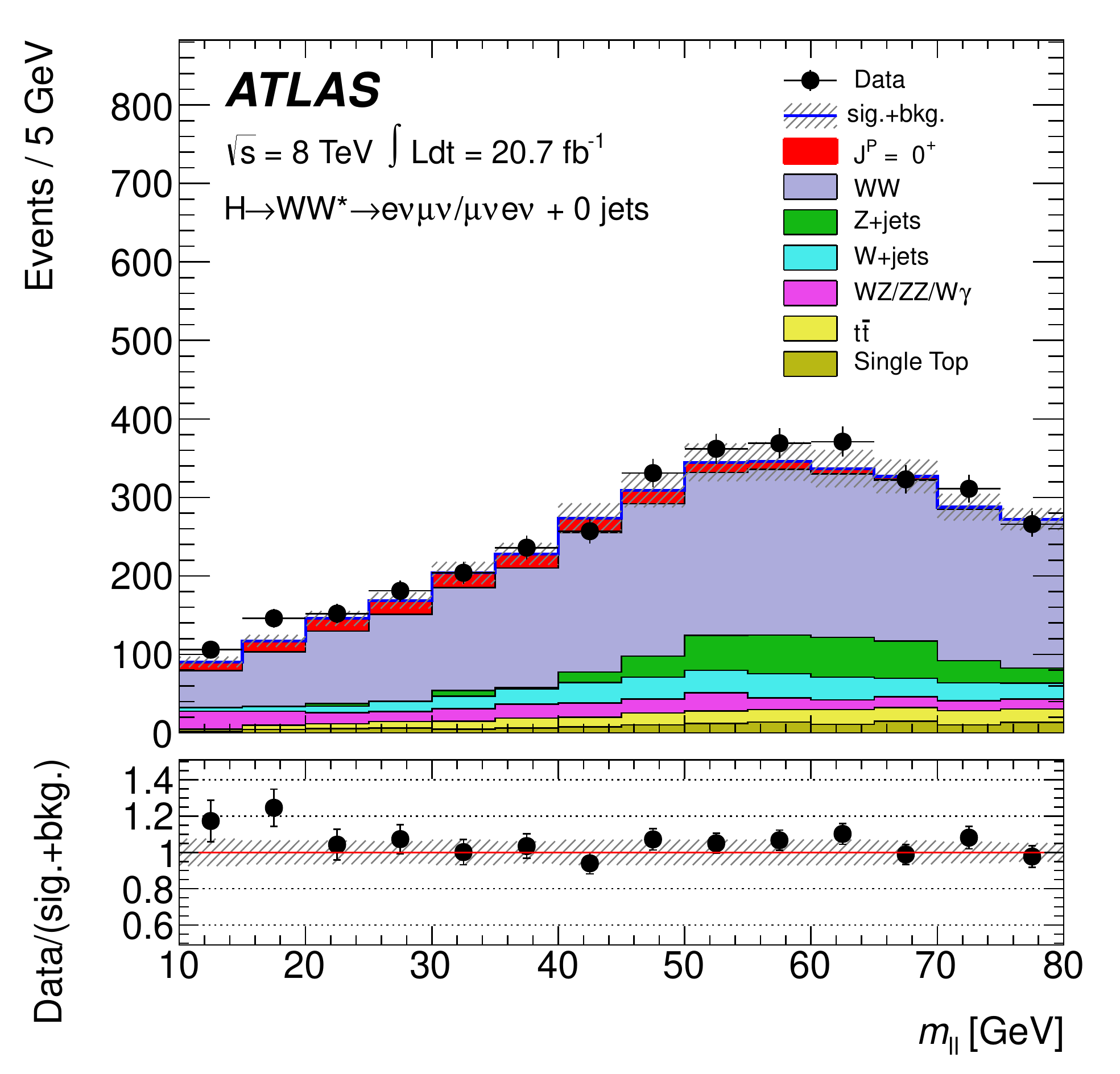}}
\caption{Distributions of (a) \dphill\ and (b) \mll\ in the signal region for $m_{H} = 125\GeV$ and the \spinzero\ hypothesis. 
    The signal is normalised to its SM expectation. In the lower part of the figures the ratio between data and the sum of signal and background is shown.
   The hatched areas represent the uncertainty on the signal and background yields from statistical, experimental, and theoretical sources.
  }
\label{fig:SRInputV}
\end{figure}

Spin correlations between the decay products affect the $\hwwlnln$ event topologies by shaping the angular
distributions of the leptons as well as the distributions of the lepton momenta and missing transverse energy. 
Due to the presence of two neutrinos in the event, a direct calculation of the various decay angles is not possible. 
Two of the most sensitive variables for measuring the spin of the Higgs boson are the dilepton invariant mass, \mll, and the azimuthal separation 
of the two leptons, \dphill. 
Figure~\ref{fig:SRInputV} shows the distributions 
of both variables in the signal region. The distributions observed in the data agree
well with the MC prediction for the expected SM \spinzero\ signal.
The dilepton transverse momentum, \ptll , also has sensitivity to different spin hypotheses.

A BDT algorithm is used to distinguish between the spin hypotheses. In addition to the three variables 
mentioned above, the transverse mass of the dilepton and missing momentum system, \mT ~\cite{ATLAScouplings}, is used in 
the BDT training as it provides a good separation between backgrounds and signals as well as some separation between 
the different spin hypotheses for the signals. Two separate BDT classifiers are developed for each hypothesis test: 
one classifier is trained to distinguish the \spinzero\ signal 
from the sum of all backgrounds while the second classifier separates the alternative spin--parity hypothesis 
(\spintwo, \spinonep\ or \spinonem) from the sum of all backgrounds.  
Background processes used to train both
classifiers include $WW$, $t\bar{t}$ and single top, as well as $WZ$,
$ZZ$, $W\gamma$, $W\gamma^*$, $W$+jets and \Zjets .  

\begin{figure}[!htbp]
\centering
\subfigure[]{\includegraphics[width=1.0\columnwidth]{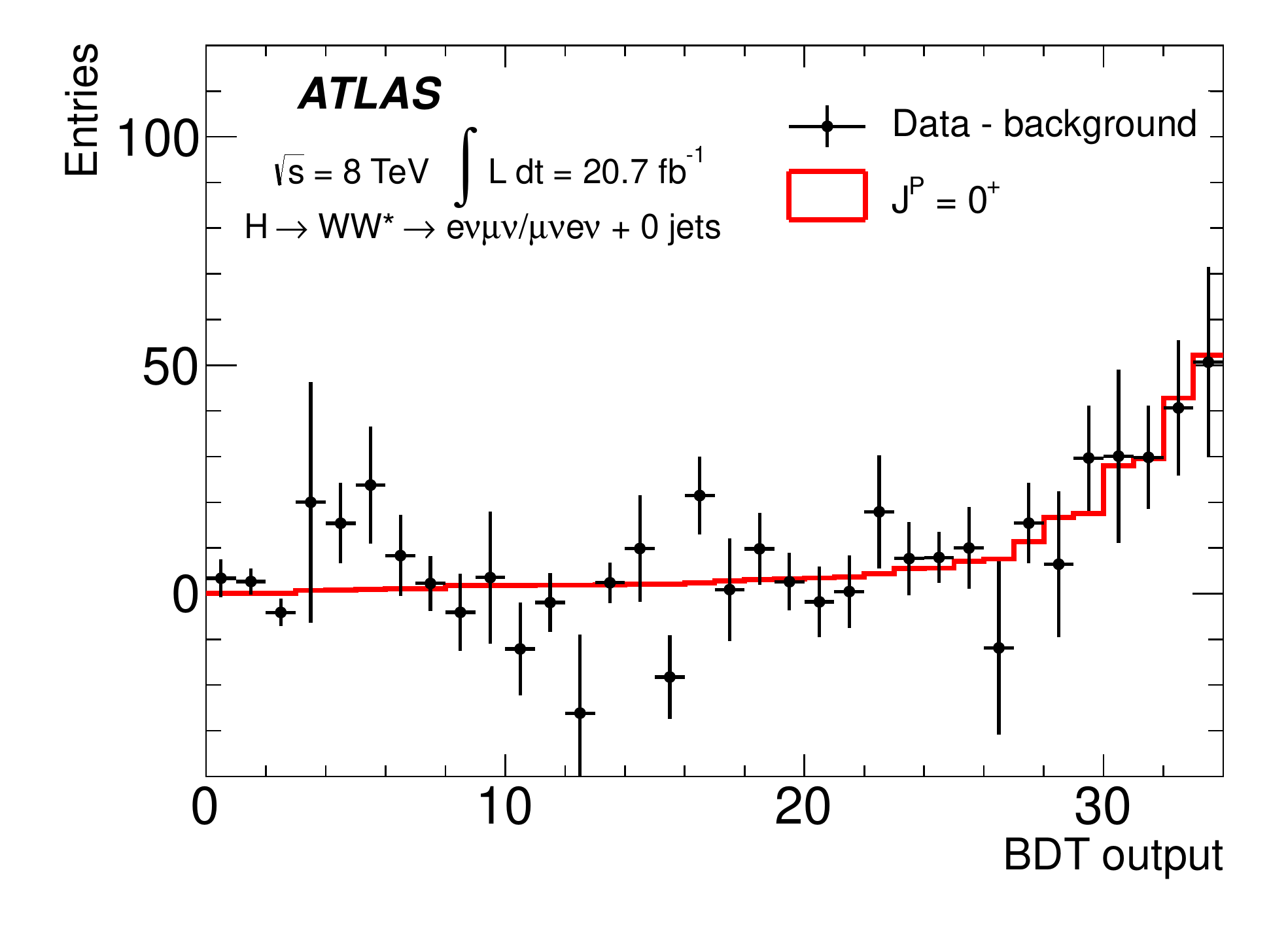}}
\subfigure[]{\includegraphics[width=1.0\columnwidth]{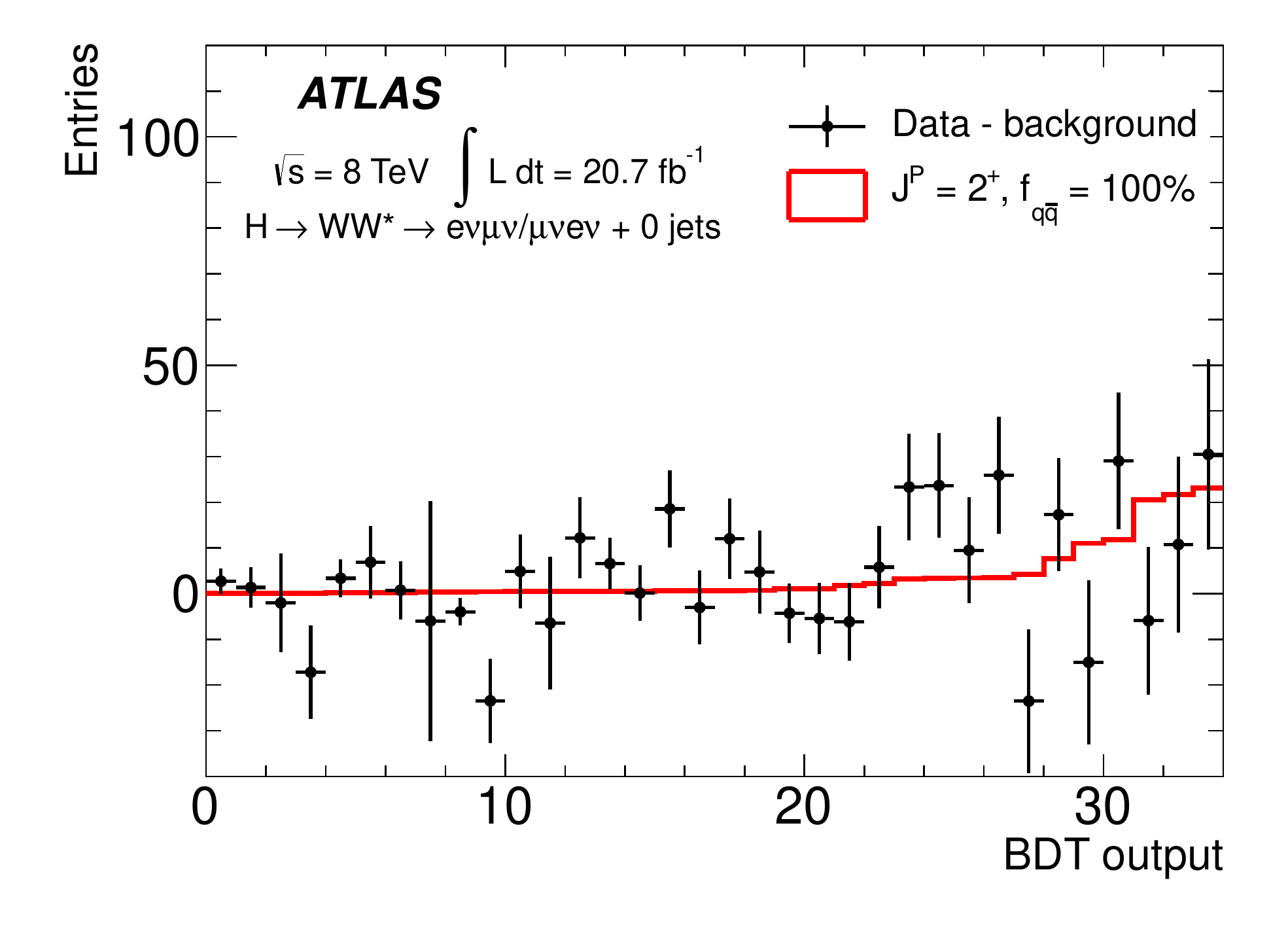}}
\caption{\label{fig:bgrSubstractedData_BDT2} 
  One-dimensional distributions of the outputs of the BDT for the \hww\ channel after background subtraction, using best-fit values for (a)
  \spinzero\ and (b) \spintwo\ with \fqqf{100} hypotheses. In each case, the two-dimensional 
  distribution of the two classifiers is remapped into a one-dimensional distribution, with the bins reordered in increasing number 
  of expected signal events. Empty bins, defined as bins with expected content below 0.1, are removed.}
\end{figure}

The resulting two-dimensional distribution of the two classifiers is then used in binned likelihood fits to test the data for compatibility 
with the presence of a \spinzero, \spinonep, \spinonem or \spintwop\ particle in the data. The analysis of \spintwo, including 
the retraining of the second classifier with the \spintwo\ sample as signal, is repeated for each of the five values of \fqq.
The BDT output distributions for data, after background subtraction, are shown in Fig.~\ref{fig:bgrSubstractedData_BDT2}, 
after remapping the two-dimensional distribution of the two classifiers into a one-dimensional distribution.

The BDT relies on a good description of the input variables and their correlations.
These were studied in detail and found to be well described by simulation~\cite{conf_ww}. In addition, dedicated 
studies were performed to verify that a BDT with the chosen four 
input variables is able to reliably separate the main backgrounds in a 
background-enriched region, and that the response is well modelled. 

Two different categories of systematic uncertainties are considered: experimental or
detector sources, such as the jet energy scale and resolution, or the lepton identification efficiencies, scale and resolution, 
as well as theoretical sources  such as the estimation of the effect of higher-order contributions 
through variations of the QCD renormalisation and factorisation scales in the Monte Carlo simulation.  The experimental 
uncertainties affect both the signal and background yields and are described in 
Ref.~\cite{ATLAScouplings}. Monte Carlo  samples with systematically varied parameters 
were analysed. 
Both the overall normalisation and shape distortions
are included as nuisance parameters in the likelihood. 

The $WW$ background in the signal region is evaluated through extrapolation from 
a control region using the simulation. The theoretical uncertainties on the extrapolation
parameter $\alpha = N_{\mathrm SR}/N_{\mathrm CR}$, the ratio of the number of events passing 
the signal region selection to the number passing the control region selection, are evaluated according 
to the prescription of Ref.~\cite{Dittmaier:2012vm}. Several sources of uncertainty on the normalisation are considered: 
uncertainties on the QCD renormalisation and factorisation scales, Parton Density Functions (PDF), the choice of Monte Carlo generator, 
and the underlying event and parton shower model.
The total uncertainty on the extrapolation is $\pm 4.8$\%. Another important uncertainty 
arises from the shape modelling of the irreducible $WW$ continuum background. The uncertainty on the shapes
of the BDT discriminants is studied by varying the factorisation and renormalisation scales, by comparing the predictions of HERWIG~\cite{herwig} and 
 PYTHIA8 leading-order parton shower programs, and by evaluating the uncertainties from the CT10~\cite{Lai:2010vv} PDF error set and combining them with the 
difference in central values between NNPDF~\cite{Ball:2011mu} and CT10. An envelope for the predicted BDT shape for each discriminant is derived and 
included in the binned likelihood fit as a shape uncertainty.

\section{Results} 
\label{sec:Results}
The SM \spinzero\ hypothesis is tested against several alternative spin--parity hypotheses using 
the analyses described in the previous sections.
Using the statistical procedure described in Section~\ref{sec:Statistics}, 
integral probabilities, the $p_0$-values, are determined to quantify the level of agreement of 
the data with different spin--parity hypotheses. When giving the confidence level associated with the rejection of 
a spin--parity hypothesis, the \CLs\ approach defined in Eq.~\ref{eq:cls} is used.

\subsection{Systematic uncertainties}
\label{sec:systematics}
The sources of systematic uncertainty accounted for in the analyses of the individual channels are 
discussed in Sections~\ref{sec:hgg}, ~\ref{sec:hzz} and ~\ref{sec:hww}. 
In the combination, the correlations among the common sources of systematic uncertainties across channels are taken into account. 
Systematic uncertainties on electron and muon identification, reconstruction and trigger efficiencies, as well as on their energy and momentum resolution,
 are common to both the \hzz\ and \hww\ channels.
Systematic uncertainties on the energy scale of electrons and photons affect all three  decay channels. 
It was also verified that the results presented in the following are insensitive to variations of the Higgs boson mass within the measured 
accuracy of about $\pm 0.6$ GeV~\cite{ATLAScouplings}.

The overall impact of the systematic uncertainties is evaluated by comparing the baseline results of the likelihood fits obtained by profiling 
all nuisance parameters not directly measured from the data, with the results obtained by 
fixing them at their nominal values.
For all tested hypotheses, the combined rejection significance is found to 
be degraded by less than 0.3$\sigma$ when including all nuisance parameters in the fit with respect to 
fixing them at their nominal values.

The production mode has a significant impact on the underlying \pt\ spectrum of the Higgs boson. 
For signals produced through gluon fusion, the dependence on the \pt\ modelling was studied by comparing 
the discriminant observables before and after re-weighting the signal to the POWHEG+PYTHIA8 spectrum.
However, the impact on the discriminant observables is found to be negligible compared to other sources of systematic uncertainty and 
therefore is neglected.
For the $q \bar{q}$-initiated processes the \pt\ spectrum is expected to be softer than for processes produced via gluon fusion.
Since no higher-order QCD predictions are available for the $q \bar{q}$ annihilation production process, no specific systematic 
uncertainty is assigned to the \pt\ spectrum of such signals. 
The impact of the large variation obtained by re-weighting the signals produced at leading order in $q \bar{q}$ annihilation for the \spintwo\ model to the 
POHWEG+PYTHIA8 gluon-fusion prediction was evaluated. 
The resulting weights increase from about unity at low transverse momentum to about four near 100 GeV. 
The \hww\ and \hzz\ channels are almost insensitive to such re-weighting, which leads to changes in the BDT discriminant shapes of the order of a few percent.
The \hgg\ channel is more sensitive to the signal \pt\ spectrum due to the impact on its acceptance at high \costs\ values. 
For this channel, the expected sensitivity for 
the spin-2 rejection is reduced by about 30\%\  for $f_{q \bar{q}} = 100 \%$, when the re-weighting is applied.
Since the combined result for this case is dominated by the \hzz\ and \hww\ channels, the overall impact of this 
re-weighting on the combined \spintwo\ rejection is negligible, below 0.1$\sigma$.

\subsection{Test of SM $J^P = 0^+$ against $J^P = 0^-$  \label{sec:zerom}}
\begin{figure}[!htbp]
  \centering
  \includegraphics[width=1.0\columnwidth]{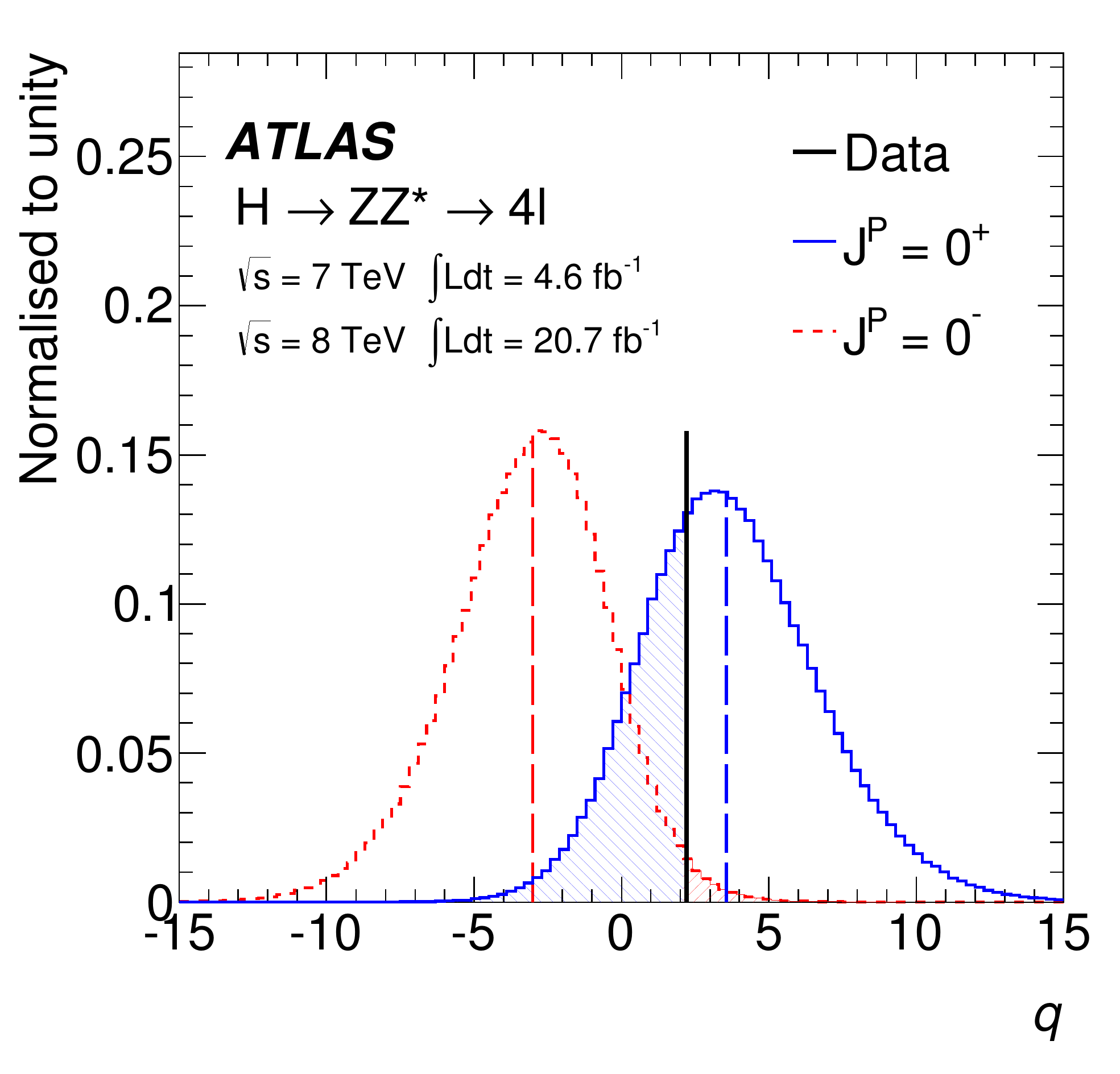}
  \caption{Expected distributions of $q = \log({\cal L} (\spinzero )/{\cal L}( \spinzeromm ))$, the logarithm of the ratio of profiled likelihoods, under 
  the \spinzero\ and \spinzerom\ hypotheses for the Standard Model  \spinzero\ (blue/solid line distribution) or  \spinzerom\ (red/dashed line distribution) 
  signals.  The observed value is indicated by the vertical solid line and the expected medians by the dashed lines. 
  The coloured areas correspond to the integrals of the expected distributions up to the 
 observed value and are used to compute the $p_0$-values for the rejection of each hypothesis. }
  \label{fig:ZZ_LLR}
\end{figure}

The distributions of the test statistics $q$ from the \hzz\ channel for the  \spinzero\  and \spinzerom\  hypotheses 
are shown in Fig.~\ref{fig:ZZ_LLR} together with the observed value.  

The expected and observed rejections of the \spinzero\ and \spinzerom\ hypotheses 
are summarised in Table~\ref{tab:resultsummary_0p0m}. The data are in agreement with the \spinzero\ 
hypothesis, while the \spinzerom\ hypothesis is excluded at 97.8\% CL. 

\subsection{Test of SM $J^P = 0^+$ against $J^P = 1^+$  \label{sec:onep}}
The expected and observed rejections of the \spinzero\ and \spinonep\ hypotheses in the \hzz\ and \hww\ channels and their
combination are summarised in Table~\ref{tab:resultsummary_0p1p}. For both channels, the results are in agreement with the \spinzero\ 
hypothesis. In the \hzz\ channel, the \spinonep\ hypothesis is excluded at  99.8\% CL, 
while in the \hww\ channel, it is excluded at  92\% CL. 
The combination excludes this hypothesis at 99.97\% CL. 

\subsection{Test of SM $J^P = 0^+$ against $J^P = 1^-$  \label{sec:onem}}
The expected and observed rejections of the \spinzero\ and \spinonem\ hypotheses in the \hzz\ and \hww\ channels and their
combination are summarised in Table~\ref{tab:resultsummary_0p1m}. For both channels, the results are in agreement with the \spinzero\ 
hypothesis. In the \hzz\ channel, the \spinonem\ hypothesis is excluded at  94\% CL.
In the \hww\ channel, the \spinonem\ hypothesis is excluded at 98\% CL. The combination excludes this hypothesis at~99.7\%~CL. 

\subsection{Test of SM $J^P = 0^+$ against $J^P = 2^+$  \label{sec:twop}}
The expected and observed rejections of the \spinzero\ and  $2^{+}$ hypotheses in the three channels are summarised in 
Table~\ref{tab:resultsummary_combined}, for  all  \fqq\ values of the spin-2 particle considered. 
For all three channels, the results are in agreement with the spin-0 hypothesis. 
The results from the \hgg\ channel exclude a spin-2 particle produced via gluon fusion ($\fqq  =0$) at 99.3\% CL. 
The separation between the two spin hypotheses in this channel decreases with increasing \fqq.  
For large values of \fqq , the \costs\ distributions associated with the spin-0 and spin-2 signals become very similar.
In the case of the \hzz\ channel, a separation slightly above one standard deviation is expected 
between the \spinzero\ and \spintwo\ hypotheses, with little dependence on the production mechanism. 
The \hww\ channel has the opposite behaviour to the \hgg\ one, with the best expected rejection achieved for large 
values of \fqq , as illustrated in Table~\ref{tab:resultsummary_combined}.
The results for the \hww\ channel are also in agreement with the \spinzero\ hypothesis. 
The \spintwo\ hypothesis is excluded with a CL above 95\%. 
The data are in better agreement with the \spinzero\ hypothesis over the full range of \fqq .

\begin{figure}[!htbp]
  \centering
  \includegraphics[width=1.0\columnwidth]{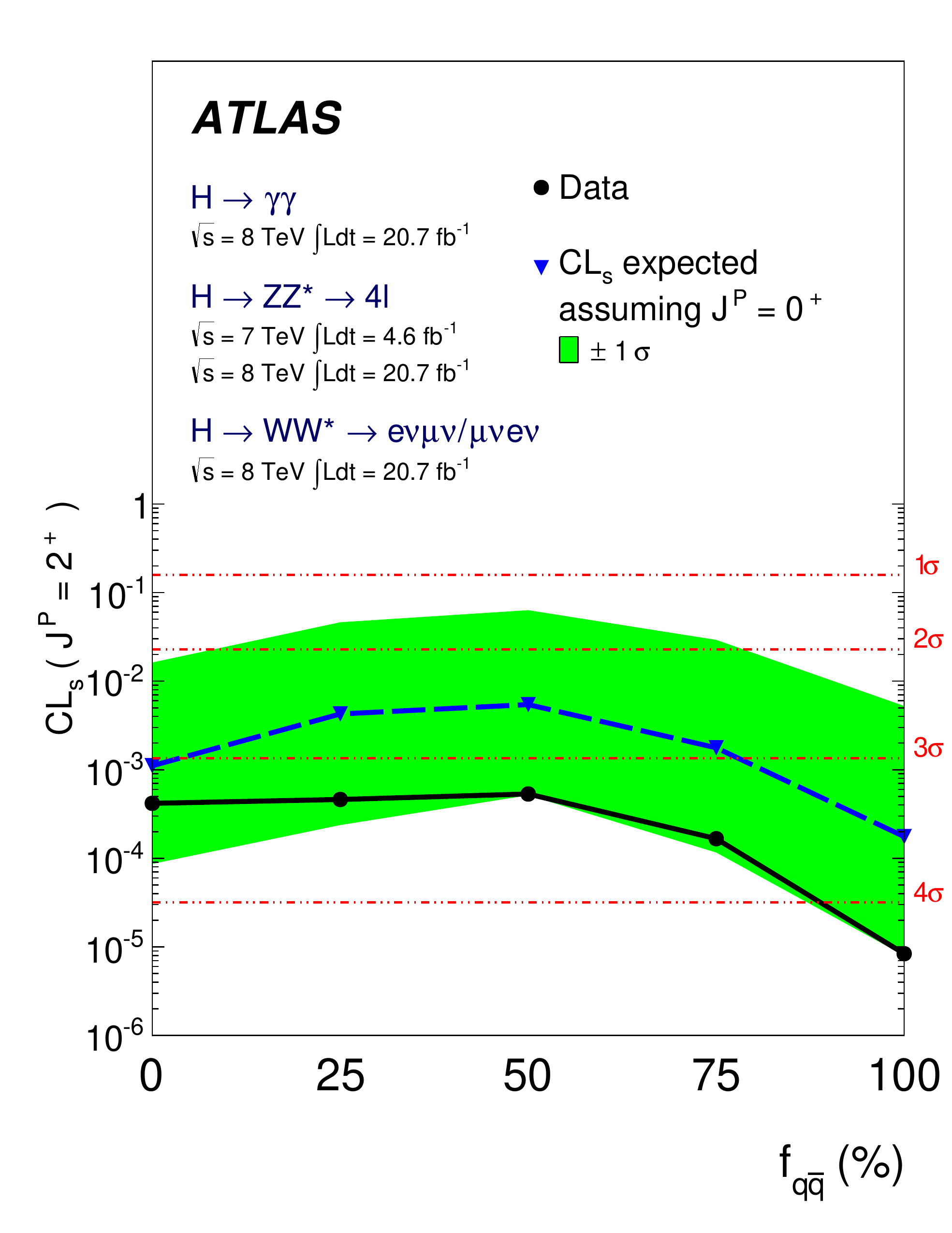}
  \caption{Expected (blue triangles/dashed line) and observed (black circles/solid line) confidence levels, $\CLs(\spintwo)$, of the \spintwo\
 hypothesis as a function of the fraction \fqq\ (see text) for the spin-2 particle. 
 The green bands represent the 68\%\ expected exclusion range for a signal with assumed \spinzero .
  On the right $y$-axis, the corresponding numbers of Gaussian standard deviations are given, using the one-sided convention.}
  \label{fig:CLs}
\end{figure}

Table~\ref{tab:res1} shows the expected and observed $p_0$-values for both the \spinzero\ and \spintwo\ hypotheses for the combination of 
the \hgg, \hzz\ and \hww\ channels. 
The test statistics calculated on data are compared to the corresponding expectations obtained from 
pseudo-experiments, as a function of \fqq. 
The data are in good agreement with the Standard Model \spinzero\ hypothesis over the full \fqq\ range.
Figure~\ref{fig:CLs} shows the comparison of the expected and observed \CLs\ values for the \spintwo\ rejection as a function of \fqq. 
The observed exclusion of the \spintwo\ hypothesis in favour of the Standard Model
\spinzero\ hypothesis exceeds 99.9\%~CL for all values of \fqq.

\begin{figure}[!htbp]
  \centering
  \includegraphics[width=1.0\columnwidth]{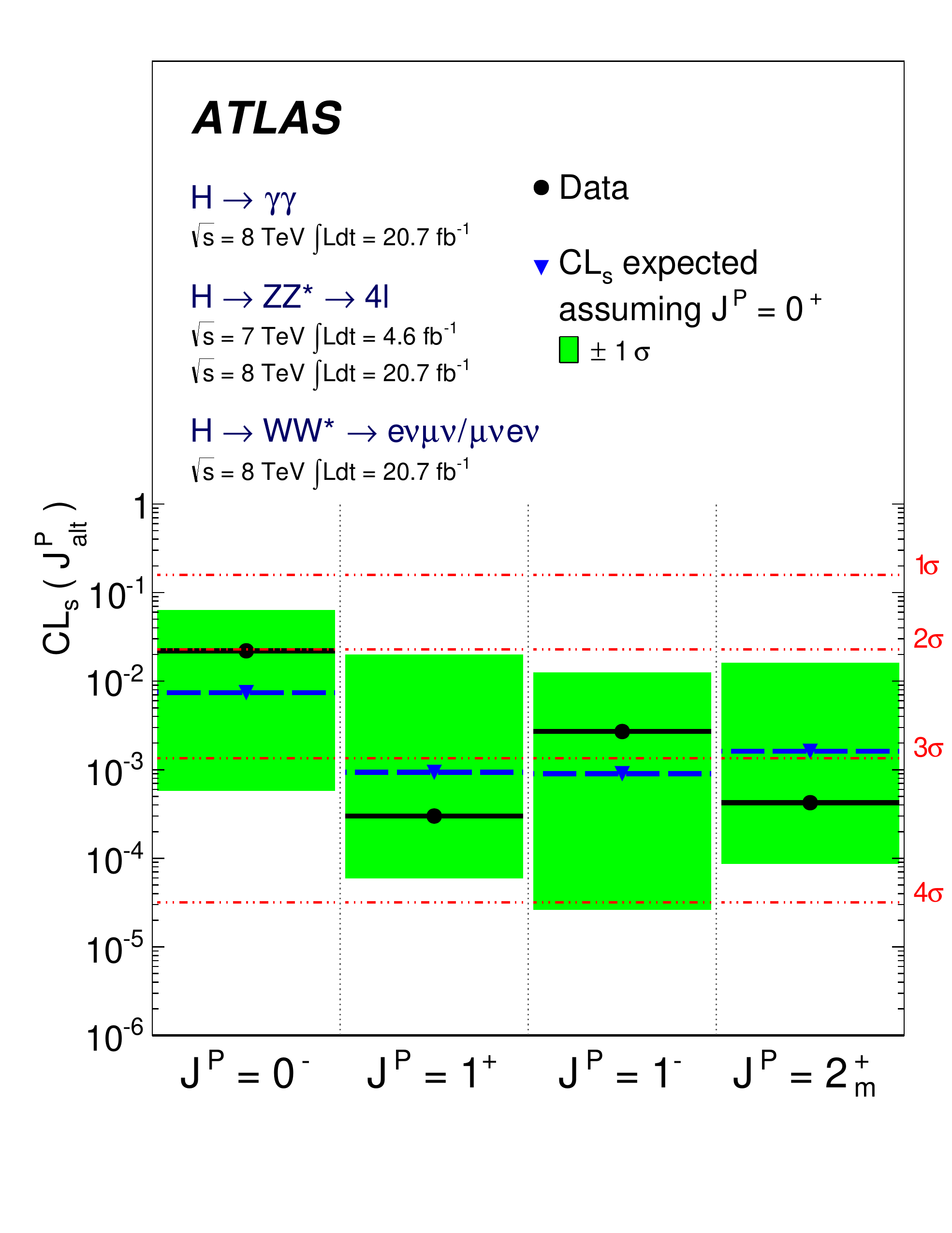}
  \caption{Expected (blue triangles/dashed lines) and observed (black circles/solid lines) confidence level $\CLs$ for alternative spin--parity hypotheses 
  assuming  a \spinzero\ signal.  
  The green band represents the 68\%\  $\CLs(\spinalt)$ expected exclusion range for a signal with assumed \spinzero .
  For the spin-2 hypothesis, the results for the specific \spintwopm\ model, discussed in Section~\ref{sec:SpinMC}, are shown. 
  On the right $y$-axis, the corresponding numbers of Gaussian standard deviations are given, using the one-sided convention.}
  \label{fig:summary}
\end{figure}

\subsection{Summary}
The observed and expected \CLs\ values for the exclusion of the different spin--parity hypotheses 
are summarised in Fig.~\ref{fig:summary}.  For the spin-2 hypothesis, the \CLs\ value for the specific \spintwopm\ model, discussed 
in Section~\ref{sec:SpinMC}, is displayed.

\section{Conclusions}
\label{sec:Conclusion}
The Standard Model \spinzero\ hypothesis for the Higgs boson has been compared to alternative spin--parity hypotheses using  
$\sqrt{s} = 8$ TeV (20.7 \ifb) and 7 TeV (4.6 \ifb)
proton--proton collision data collected by the ATLAS experiment at the LHC.
The Higgs boson decays \hgg , \hzzllll\ and \hwwlnln\ have been used to test several specific alternative models, 
including $J^{P} = 0^{-}, 1^{+}, 1^{-}$ and a graviton-inspired \spintwo\ model with minimal couplings to SM particles.
The data favour the Standard Model quantum numbers of \spinzero .
The \spinzerom\ hypothesis is rejected at 97.8\%\ CL by using the \hzzllll\ decay alone. 
The \spinonep\ and \spinonem\ hypotheses are rejected with a CL of at least 99.7\%\ by combining the \hzzllll\ and \hwwlnln\ channels.
Finally, the \spintwo\ model is rejected at more than 99.9\%\ CL by combining all three bosonic channels, \hgg , \hzzllll\ and \hwwlnln , 
independent of the assumed admixture of gluon-fusion and quark--antiquark production.
All alternative models studied in this Letter are excluded without assumptions on the strength of 
the couplings of the Higgs boson to SM particles.
These studies provide evidence for the spin-0 nature of the Higgs boson, with positive parity being strongly preferred.

\section*{Acknowledgements}
We thank CERN for the very successful operation of the LHC, as well as the
support staff from our institutions without whom ATLAS could not be
operated efficiently.

We acknowledge the support of ANPCyT, Argentina; YerPhI, Armenia; ARC,
Australia; BMWF and FWF, Austria; ANAS, Azerbaijan; SSTC, Belarus; CNPq and FAPESP,
Brazil; NSERC, NRC and CFI, Canada; CERN; CONICYT, Chile; CAS, MOST and NSFC,
China; COLCIENCIAS, Colombia; MSMT CR, MPO CR and VSC CR, Czech Republic;
DNRF, DNSRC and Lundbeck Foundation, Denmark; EPLANET, ERC and NSRF, European Union;
IN2P3-CNRS, CEA-DSM/IRFU, France; GNSF, Georgia; BMBF, DFG, HGF, MPG and AvH
Foundation, Germany; GSRT and NSRF, Greece; ISF, MINERVA, GIF, DIP and Benoziyo Center,
Israel; INFN, Italy; MEXT and JSPS, Japan; CNRST, Morocco; FOM and NWO,
Netherlands; BRF and RCN, Norway; MNiSW, Poland; GRICES and FCT, Portugal; MERYS
(MECTS), Romania; MES of Russia and ROSATOM, Russian Federation; JINR; MSTD,
Serbia; MSSR, Slovakia; ARRS and MIZ\v{S}, Slovenia; DST/NRF, South Africa;
MICINN, Spain; SRC and Wallenberg Foundation, Sweden; SER, SNSF and Cantons of
Bern and Geneva, Switzerland; NSC, Taiwan; TAEK, Turkey; STFC, the Royal
Society and Leverhulme Trust, United Kingdom; DOE and NSF, United States of
America.

The crucial computing support from all WLCG partners is acknowledged
gratefully, in particular from CERN and the ATLAS Tier-1 facilities at
TRIUMF (Canada), NDGF (Denmark, Norway, Sweden), CC-IN2P3 (France),
KIT/GridKA (Germany), INFN-CNAF (Italy), NL-T1 (Netherlands), PIC (Spain),
ASGC (Taiwan), RAL (UK) and BNL (USA) and in the Tier-2 facilities
worldwide.

\bibliographystyle{atlasBibStyleWithTitle} 
\bibliography{main} 

\clearpage
\begin{table*}[hbtb]
\caption{\label{tab:resultsummary_0p0m} Summary of results for the \spinzerop\ versus \spinzerom\ test in the  \hzz\  channel. 
The expected $p_0$-values for rejecting the \spinzerop\ and \spinzerom\ hypotheses (assuming the alternative hypothesis) are shown in the second and third columns. 
The fourth and fifth columns show the observed $p_0$-values, while the \CLs\ value for excluding the \spinzerom\ hypothesis is given in the last column.}
\vspace{1mm}
\renewcommand{\arraystretch}{1.0}
\centering
\renewcommand{\arraystretch}{1.05}
\begin{tabular}{c||c|c||c|c||c}
\hline\hline
\multirow{2}{*}{Channel} & $0^{-}$ assumed & $0^{+}$ assumed & \multirow{2}{*}{Obs. $p_0(\spinzero)$} & \multirow{2}{*}{Obs.
$p_0(\ensuremath{J^P=0^-})$} & \multirow{2}{*}{$\CLs (\ensuremath{J^P=0^-})$} \\
   & Exp. $p_0(\spinzero)$ & Exp. $p_0(J^P=0^-)$ & & & \\
\hline
 \hzz\ &       $1.5 \cdot 10^{-3}$ & $3.7 \cdot 10^{-3}$ & 0.31 & 0.015              & 0.022 \\
\hline\hline
\end{tabular}
\end{table*}

\begin{table*}[htbp]
\caption{\label{tab:resultsummary_0p1p} Summary of results for the \spinzero\ versus \spinonep\ test in the \hzz\ and \hww\  channels,
as well as their combination. The expected $p_0$-values for rejecting the \spinzero\ and
\spinonep\ hypotheses (assuming the alternative hypothesis) are shown in the second and third columns. The fourth and fifth
columns show the observed $p_0$-values, while the \CLs\ values for excluding the \spinonep\ hypothesis are given in the last column.}
\vspace{1mm}
\renewcommand{\arraystretch}{1.0}
\centering
\renewcommand{\arraystretch}{1.05}
\begin{tabular}{c||c|c||c|c||c}
\hline\hline
\multirow{2}{*}{Channel} & $1^{+}$ assumed & $0^{+}$ assumed & \multirow{2}{*}{Obs. $p_0(\spinzero)$} & \multirow{2}{*}{Obs.
$p_0(\ensuremath{J^P=1^+})$} & \multirow{2}{*}{$\CLs (\ensuremath{J^P=1^+})$} \\
   & Exp. $p_0(\spinzero)$ & Exp. $p_0(\ensuremath{J^P=1^+})$ & & & \\
\hline
 \hzz\ &       $4.6 \cdot 10^{-3}$ & $1.6 \cdot 10^{-3}$ & 0.55 & $1.0 \cdot 10^{-3}$ & $2.0 \cdot 10^{-3}$ \\
 \hww\ &       0.11                & 0.08                & 0.70 & 0.02                & 0.08 \\
 Combination & $2.7 \cdot 10^{-3}$ & $4.7 \cdot 10^{-4}$ & 0.62 & $1.2 \cdot 10^{-4}$ & $3.0 \cdot 10^{-4}$ \\
\hline\hline
\end{tabular}
\end{table*}

\begin{table*}[hbtb]
\caption{\label{tab:resultsummary_0p1m} Summary of results for the \spinzero\ versus \spinonem\ test in the \hzz\ and \hww\  channels,
as well as their combination. The expected $p_0$-values for rejecting the \spinzero\ and
\spinonem\ hypotheses (assuming the alternative hypothesis) are shown in the second and third columns. The fourth and fifth
columns show the observed $p_0$-values, while the $\CLs$ values for excluding the \spinonem\ hypothesis are given in the last column.}
\vspace{1mm}
\renewcommand{\arraystretch}{1.0}
\centering
\renewcommand{\arraystretch}{1.05}
\begin{tabular}{c||c|c||c|c||c}
\hline\hline
\multirow{2}{*}{Channel} & $1^{-}$ assumed & $0^{+}$ assumed & \multirow{2}{*}{Obs. $p_0(\spinzero)$} & \multirow{2}{*}{Obs.
$p_0(\ensuremath{J^P=1^-})$} & \multirow{2}{*}{$\CLs (\ensuremath{J^P=1^-})$} \\
   & Exp. $p_0(\spinzero)$ & Exp. $p_0(\ensuremath{J^P=1^-})$ & & & \\
\hline
 \hzz\ &       $0.9 \cdot 10^{-3}$ & $3.8 \cdot 10^{-3}$ & 0.15 & 0.051               & 0.060 \\
 \hww\ &       0.06                & 0.02                & 0.66 & 0.006               & 0.017 \\
 Combination & $1.4 \cdot 10^{-3}$ & $3.6 \cdot 10^{-4}$ & 0.33 & $1.8 \cdot 10^{-3}$ & $2.7 \cdot 10^{-3}$ \\
\hline\hline
\end{tabular}
\end{table*}

\clearpage 

\begin{table*}[hbtp]
\caption{\label{tab:resultsummary_combined} Summary of results for the various fractions \fqq\ of the \qqbar\ production of the spin-2 particle for
the \hgg\ (top), \hzz\ (middle), and \hww\ (bottom)  channels. The expected $p_0$-values for rejecting the \spinzero\ and \spintwo\
hypotheses (assuming the alternative hypothesis) are shown in the second and third columns. The fourth and fifth columns show the observed
$p_0$-values, while the \CLs values for excluding the \spintwo\ hypothesis are given in the last column.}
\vspace{1mm}
\centering
\renewcommand{\arraystretch}{1.05}
\resizebox{\textwidth}{!}{
\begin{tabular}{c||c|c||c|c||c}
\multicolumn{6}{c}{\hgg}\\
\hline\hline
\multirow{2}{*}{\fqq} & $2^{+}$ assumed & $0^{+}$ assumed & \multirow{2}{*}{Obs. $p_0(\spinzero)$} & \multirow{2}{*}{Obs. $p_0(\spintwo)$} & \multirow{2}{*}{$\CLs (\spintwo)$} \\
   & Exp. $p_0(\spinzero)$ & Exp. $p_0(\spintwo)$ & & & \\
\hline
100\% & 0.148 & 0.135 & 0.798 & 0.025 & 0.124 \\
 75\% & 0.319  & 0.305 & 0.902 & 0.033 & 0.337 \\
 50\% &  0.198 & 0.187 & 0.708 & 0.076 & 0.260 \\
 25\% & 0.052 & 0.039 & 0.609 & 0.021 & 0.054 \\
 0\% &  0.012  & 0.005 & 0.588 & 0.003 & 0.007 \\
\hline\hline
\multicolumn{6}{c}{\ }\\
\multicolumn{6}{c}{\hzz}\\
\hline\hline
\multirow{2}{*}{\fqq} &  $2^{+}$ assumed assumed &  $0^{+}$ assumed & \multirow{2}{*}{Obs. $p_0(\spinzero)$} & \multirow{2}{*}{Obs. $p_0(\spintwo)$} & \multirow{2}{*}{$\CLs (\spintwo)$} \\
   & Exp. $p_0(\spinzero)$ & Exp. $p_0(\spintwo)$ & & & \\
\hline
100\% & 0.102  &  0.082  &  0.962 & 0.001&  0.026\\
 75\%  &   0.117 &  0.099 &   0.923& 0.003&  0.039\\
 50\%  &   0.129 &  0.113 &   0.943& 0.002& 0.035\\
 25\%  &   0.125 &  0.107 &    0.944& 0.002& 0.036\\
 0\%    &  0.099  &   0.092 &  0.532 & 0.079& 0.169\\
\hline\hline
\multicolumn{6}{c}{\ }\\
\multicolumn{6}{c}{\hww}\\
\hline\hline
\multirow{2}{*}{\fqq} &  $2^{+}$ assumed &  $0^{+}$ assumed & \multirow{2}{*}{Obs. $p_0(\spinzero)$} & \multirow{2}{*}{Obs. $p_0(\spintwo)$} & \multirow{2}{*}{$\CLs (\spintwo)$} \\
   & Exp. $p_0(\spinzero)$ & Exp. $p_0(\spintwo)$ & & & \\
\hline
100\% & 0.013 & $3.6\cdot 10^{-4}$ & 0.541 & $1.7\cdot 10^{-4}$ & $3.6\cdot 10^{-4}$\\
    75\%  & 0.028 & 0.003 & 0.586 & 0.001 &  0.003\\
    50\% &   0.042 & 0.009 & 0.616 & 0.003 & 0.008\\
    25\% &   0.048 & 0.019 & 0.622 & 0.008 & 0.020\\
    0\% &  0.086 & 0.054 & 0.731 & 0.013 & 0.048 \\
    \hline\hline
\end{tabular}}
\renewcommand{\arraystretch}{1.0}
\end{table*}

\begin{table*}[htbp]
\caption{\label{tab:res1}Expected and observed $p_0$-values for the \spinzero\ and \spintwo\ hypotheses as a function of the fraction \fqq\ of the 
$q\bar{q}$ spin-2 production mechanism. The values are tabulated for the combination of the \hgg, \hzz\ and \hww\ channels. The \CLs\ values
 for excluding the \spintwo\ hypothesis are given in the last column.}
\vspace{1mm}
\renewcommand{\arraystretch}{1.0}
\centering
\renewcommand{\arraystretch}{1.05}
\resizebox{\textwidth}{!}{
\begin{tabular}{c||c|c||c|c||c}
\hline\hline
\multirow{2}{*}{\fqq} &  $2^{+}$ assumed &  $0^{+}$ assumed & \multirow{2}{*}{Obs. $p_0(\spinzero)$} & \multirow{2}{*}{Obs. $p_0(\spintwo)$} & \multirow{2}{*}{$\CLs (\spintwo)$} \\
   & Exp. $p_0(\spinzero)$ & Exp. $p_0(\spintwo)$ & & & \\
\hline
100\% & $3.0 \cdot 10^{-3}$ & $8.8 \cdot 10^{-5}$ & 0.81 & $1.6 \cdot 10^{-6}$ & $0.8 \cdot 10^{-5}$ \\ 
75\% & $9.5 \cdot 10^{-3}$ & $8.8 \cdot 10^{-4}$ & 0.81 & $3.2 \cdot 10^{-5}$ & $1.7 \cdot 10^{-4}$  \\ 
50\% & $1.3 \cdot 10^{-2}$ & $2.7 \cdot 10^{-3}$ & 0.84 & $8.6 \cdot 10^{-5}$ & $5.3 \cdot 10^{-4}$  \\ 
25\% & $6.4 \cdot 10^{-3}$ & $2.1 \cdot 10^{-3}$ & 0.80 & $0.9 \cdot 10^{-4}$ & $4.6 \cdot 10^{-4}$  \\ 
0\% & $2.1 \cdot 10^{-3}$ & $5.5 \cdot 10^{-4}$ & 0.63 & $1.5 \cdot 10^{-4}$ & $4.2 \cdot 10^{-4}$   \\ 
 \hline\hline
\end{tabular}}
\end{table*}

\clearpage
\onecolumn
\begin{flushleft}
{\Large The ATLAS Collaboration}

\bigskip

G.~Aad$^{\rm 48}$,
T.~Abajyan$^{\rm 21}$,
B.~Abbott$^{\rm 112}$,
J.~Abdallah$^{\rm 12}$,
S.~Abdel~Khalek$^{\rm 116}$,
O.~Abdinov$^{\rm 11}$,
R.~Aben$^{\rm 106}$,
B.~Abi$^{\rm 113}$,
M.~Abolins$^{\rm 89}$,
O.S.~AbouZeid$^{\rm 159}$
% ...
\bigskip
\\
% ...
$^{11}$ Institute of Physics, Azerbaijan Academy of Sciences, Baku, Azerbaijan\\
$^{12}$ Institut de F{\'\i}sica d'Altes Energies and Departament de F{\'\i}sica de la Universitat Aut{\`o}noma de Barcelona, Barcelona, Spain\\
% ...
$^{21}$ Physikalisches Institut, University of Bonn, Bonn, Germany\\
% ...
$^{48}$ Fakult{\"a}t f{\"u}r Mathematik und Physik, Albert-Ludwigs-Universit{\"a}t, Freiburg, Germany\\
% ...
$^{89}$ Department of Physics and Astronomy, Michigan State University, East Lansing MI, United States of America\\
% ...
$^{106}$ Nikhef National Institute for Subatomic Physics and University of Amsterdam, Amsterdam, Netherlands\\
% ...
$^{112}$ Homer L. Dodge Department of Physics and Astronomy, University of Oklahoma, Norman OK, United States of America\\
$^{113}$ Department of Physics, Oklahoma State University, Stillwater OK, United States of America\\
% ...
$^{116}$ LAL, Universit{\'e} Paris-Sud and CNRS/IN2P3, Orsay, France\\
% ...
$^{159}$ Department of Physics, University of Toronto, Toronto ON, Canada\\
% ...
$^{*}$ Deceased
\end{flushleft}


\end{document}